\documentclass[nofootinbib, twocolumn, pra, floatfix]{revtex4}

\usepackage{amsmath}
\usepackage{amsthm}
\usepackage{amssymb}
\usepackage{graphicx}
\usepackage{color}
\usepackage{multirow}
\usepackage[makeroom]{cancel}
\usepackage{bbm}
\usepackage[hidelinks]{hyperref}
\usepackage{qcircuit}

\usepackage{float}
\floatstyle{plaintop}
\restylefloat{table}

\DeclareMathOperator{\id}{\mathbbm{1}}
\def\ket#1{\left| #1 \right\rangle}
\def\bra#1{\left\langle #1 \right|}
\newcommand{\braket}[2]{\langle #1 | #2 \rangle}
\newcommand{\beq}{\begin{equation}}
\newcommand{\eeq}{\end{equation}}
\definecolor{JM}{RGB}{4,116,149}

\definecolor{TB}{RGB}{144,177,52}
\newcommand{\tb}[1]{\textcolor{TB}{ #1}}
\definecolor{JI}{RGB}{ 176, 58, 46 }

\begin{document}

\title{Machine learning method for state preparation and gate synthesis on photonic quantum computers}

\author{Juan Miguel Arrazola}
\email{juanmiguel@xanadu.ai}
\author{Thomas R. Bromley}
\author{Josh Izaac}
\author{Casey R. Myers}
\author{Kamil Br\'adler}
\author{Nathan Killoran}
\affiliation{Xanadu, 372 Richmond Street W, Toronto, Ontario M5V 1X6, Canada}

\begin{abstract}
We show how techniques from machine learning and optimization can be used to find circuits of photonic quantum computers that perform a desired transformation between input and output states. In the simplest case of a single input state, our method discovers circuits for preparing a desired quantum state. In the more general case of several input and output relations, our method obtains circuits that reproduce the action of a target unitary transformation. We use a continuous-variable quantum neural network as the circuit architecture. The network is composed of several layers of optical gates with variable parameters that are optimized by applying automatic differentiation using the TensorFlow backend of the Strawberry Fields photonic quantum computer simulator.
We demonstrate the power and versatility of our methods by learning how to use short-depth circuits to synthesize single photons, Gottesman-Kitaev-Preskill states, NOON states, cubic phase gates, random unitaries, cross-Kerr interactions, as well as several other states and gates. We routinely obtain high fidelities above 99\% using short-depth circuits, typically consisting of a few hundred gates. The circuits are obtained automatically by simply specifying the target state or gate and running the optimization algorithm.
\end{abstract}
\maketitle

\section{Introduction}
Machine learning is a paradigm where data is used to train computer models in order to reproduce a desired behavior, without the need for explicitly programming an algorithm  \cite{kotsiantis2007supervised, michalski2013machine, lecun2015deep, schmidhuber2015deep, goodfellow2016deep}. Quantum computing is a model of computation where the fundamental operations employed to process information are determined by the laws of quantum mechanics \cite{lloyd1996universal, nielsen2002quantum, menicucci2006universal, gu2009quantum, ladd2010quantum}. Recently, a new field known as quantum machine learning has emerged, combining insights from these two disciplines \cite{schuld2014quest,wittek2014quantum, schuld2015introduction, biamonte2017quantum}. There are many aspects to quantum machine learning, including quantum speedups for machine learning algorithms \cite{harrow2009quantum,wiebe2012quantum,lloyd2014quantum,rebentrost2014quantum}, employing classical machine learning to analyze quantum systems \cite{khaneja2005optimal, hentschel2010machine, carrasquilla2017machine, carleo2017solving}, and quantum versions of models such as neural networks, Boltzmann machines, and kernel methods \cite{killoran2018continuous, wiebe2014quantum, kieferova2016tomography, rebentrost2017quantum, schuld2018circuit, amin2018quantum, schuld2018quantum, havlicek2018supervised}.

It is natural to ask whether we can employ methods from machine learning to aid in the design and realization of quantum algorithms \cite{krenn2016automated, cincio2018learning, melnikov2018active, chen2018universal,knott2016search,morales2018variationally,wan2018learning,innocenti2018supervised,gao2018experimental}. For instance, many quantum algorithms employ subroutines that require the preparation of resource states or demand a compact decomposition of specific transformations into gates from a universal set \cite{childs2010quantum,sefi2011decompose, childs2017quantum, lau2017quantum}. For state preparation, several techniques have been proposed for specific applications \cite{bimbard2010quantum,knott2016search, sanders2018black, kashiwamura2018replacing}. Providing methods to automate and optimize the construction of these subroutines would constitute a valuable and far-reaching tool for quantum computation. This is particularly important for gate synthesis because many methods rely on product-rule approximations that lead to decompositions involving a large overhead in the total number of gates \cite{trotter1959product, suzuki1993general, childs2017toward}.  

In this work, we outline how techniques from machine learning and optimization can be employed to find quantum circuits that perform a desired transformation between input and output states. In the simplest case of a single input and output, this corresponds to state preparation: finding a circuit to create a target quantum state. In its generalized form of several inputs and outputs, this is equivalent to gate synthesis: the task of obtaining a circuit that reproduces the action of a target unitary transformation. We focus on the continuous-variable (CV) model of quantum computation, which is a leading platform for optical quantum computing \cite{braunstein2005quantum, menicucci2006universal,gu2009quantum,weedbrook2012gaussian}, noting that our results can in principle extended to other models of quantum computing.

As an ansatz for the circuit architecture we employ the CV quantum neural network introduced in Ref.~\cite{killoran2018continuous}. We consider networks with a limited number of layers that perform state preparation and gate synthesis using circuits with significantly fewer gates than standard decomposition techniques. To optimize these quantum neural networks, we simulate the corresponding circuits using the Strawberry Fields software platform for photonic quantum computation \cite{killoran2018strawberry}. Strawberry Fields is equipped with a TensorFlow \cite{abadi2016tensorflow} backend which is capable of automatically computing gradients with respect to circuit parameters. These can be used to optimize the networks using TensorFlow's built-in gradient descent optimizers. We routinely obtain fidelities above 99\% using short-depth circuits consisting of roughly one hundred gates. All results are obtained automatically by just specifying the target state or gate and running the optimization algorithm, providing a significant simplification of the overall algorithm development process.\footnote{Source code for the algorithms employed in this paper is available at \url{https://github.com/XanaduAI/quantum-learning}.}

In the following, we discuss our approach for learning state preparation and gate synthesis using the CV neural network architecture. This technique is applied to find circuits that can prepare several single and two-mode quantum states. We then employ our results to find circuits capable of synthesizing various single and two-mode gates. High fidelities are observed in all cases studied, providing evidence for the wide-ranging applicability of this approach.

\section{Automated circuit design}

The problems considered in this work fall under the same general scope: given a quantum information task, find the circuit that best achieves that task. For state preparation, the goal is to prepare a particular target state starting from a fixed reference state. For gate synthesis, the task is to provide the decomposition of a desired unitary transformation into a sequence of elementary gates.
Our strategy to attack this problem is to use a variational quantum circuit approach. A variational quantum circuit is a circuit whose gates and connectivity are fixed by an ansatz, but where some or all of the gates contain free parameters. Through varying these free parameters, we can change the unitary enacted by the circuit. The two tasks listed above can be tackled with this strategy provided that (i) the variational circuit is expressive enough, and (ii) we have a well-performing method for finding good circuit parameters. To address both of these needs, we will leverage methods and ideas from classical and quantum machine learning.

\begin{figure}[t]
\input{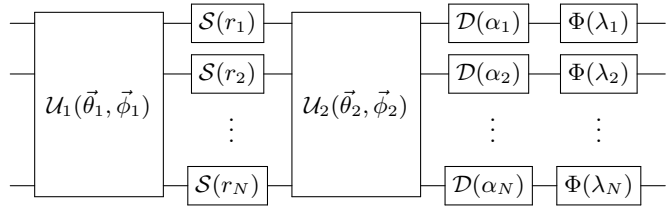}
\caption{Circuit architecture for a single layer of a continuous-variable quantum neural network. The layers consists of linear-optical interferometers $\mathcal{U}_1$, $\mathcal{U}_2$, local squeezing gates $\mathcal{S}$, local displacements $\mathcal{D}$, and non-Gaussian gates $\Phi$. }
\label{fig:layer_circuit}
\end{figure}

In order to provide the most representational power, we would like our variational circuit to be as general and flexible as possible. To this end, we will employ the CV neural network architecture from Ref. \cite{killoran2018continuous}.
The CV quantum neural network consists of multiple layers $\mathcal{L}$ composed of the sequence of gates:
\begin{equation}
\mathcal{L}:= \Phi \circ \mathcal{D} \circ \mathcal{U}_2 \circ \mathcal{S} \circ \mathcal{U}_1,
\end{equation}
where $\mathcal{U}_1=\mathcal{U}_1(\vec{\theta}_1, \vec{\phi}_1)$ and $\mathcal{U}_2=\mathcal{U}_2(\vec{\theta}_2, \vec{\phi}_2)$ are $N$-mode linear optical interferometers, $\mathcal{D}=\otimes_{i=1}^N D(\alpha_i)$ are displacements, $\mathcal{S}=\otimes_{i=1}^N S(r_i)$ are squeezing operations, and $\Phi = \otimes_{i=1}^N \phi(\lambda_i)$ are non-Gaussian gates acting independently on each mode.  The displacement and squeezing gates are defined as $D(\alpha)=\exp(\alpha a^{\dagger}-\alpha^* a)$ and $S(r)=\exp[\frac{r}{2}(a^2-{a^{\dagger}}^2)]$, where $a$ and $a^{\dagger}$ are respectively the annihilation and creation operators of the mode. The linear optical interferometers are made up of single-mode rotation gates $R(\phi)=\exp(i \phi\hat{n} )$ and two-mode beamsplitters $BS(\theta,\phi)=\exp[\theta(e^{i\phi}a_1^{\dagger}a_2-e^{-i\phi}a_1a_2^{\dagger})]$, where $\hat{n}=a^{\dagger}a$ is the number operator.  A single layer of a quantum neural network is shown in Fig. \ref{fig:layer_circuit}. Note that this network architecture is capable of simulating any universal CV quantum circuit with at most polynomial overhead, since the gates in every layer constitute a universal set \cite{lloyd1999quantum}.

The displacement, squeezing, rotation, beamsplitter, and non-Gaussian gates in a single layer all contain free parameters which determine the strength of transformation carried out by the gate. Collectively, we denote these parameters by $\vec{\theta}$. The goal is to find the choice of parameters which optimizes some cost function $C(\vec{\theta})$. In order to perform this optimization, we employ the family of gradient descent algorithms that are widely used in the field of deep learning. In vanilla gradient descent, we start off with randomly initialized values for the parameters $\vec{\theta}$. We compute the gradient $\nabla_{\vec{\theta}}\,\,C$ of the cost function with respect to the parameters $\vec{\theta}$, which determines the direction of steepest descent. We then update the parameters based on the rule
\begin{equation}
 \vec{\theta} \mapsto \vec{\theta} + \eta \nabla_{\vec{\theta}} \,\, C,
\end{equation}
where $\eta$ is some user-specified step size (also known as a learning rate). This process is iterated until the cost function no longer improves. The parameter values corresponding to the lowest-observed cost function then determine the proposed solution circuit for the given task, which may or may not be a global optimum. 

For optimization, variants of gradient descent, such as stochastic gradient descent, momentum, or Adam \cite{bottou-98x, polyak1964some, kingma2014adam} introduce noise, adaptive step-size, and memory effects into this process, yet they all follow the basic principle of following the gradient. Algorithms based on gradient descent have become so important that numerous libraries now provide the ability to automatically compute gradients \cite{bergstra2010theano,maclaurin2015autograd,abadi2016tensorflow,paszke2017automatic}. We encode the variational circuits using the Strawberry Fields software platform \cite{killoran2018strawberry} and make extensive use of its TensorFlow backend, which can automatically compute the gradients of gate parameters and perform the desired optimizations. In all cases studied in this paper, we use the Adam optimizer.

\begin{figure*}
\includegraphics[width=1.99\columnwidth]{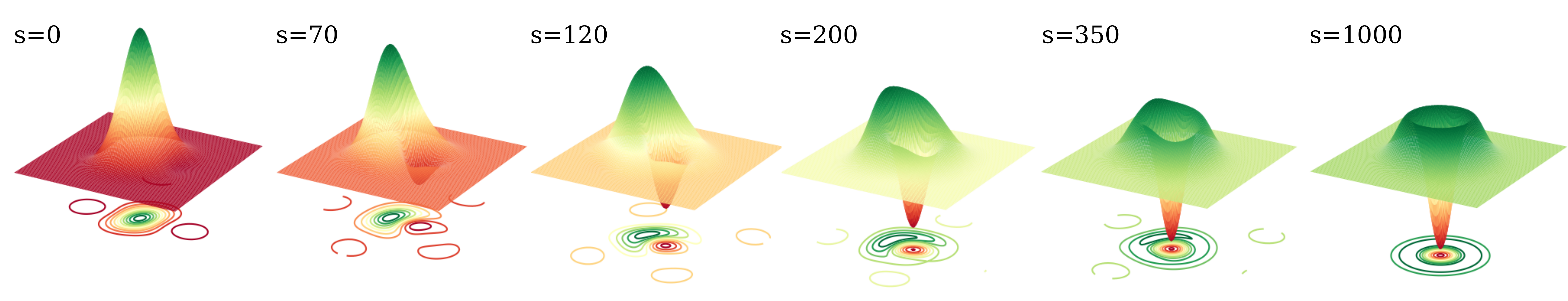}
\caption{ Wigner function of the output state of the quantum neural network during different stages of optimization performed with Strawberry Fields. The variable $s$ corresponds to the number of steps in the optimization process. Initially ($s=0$), the output state is very close to the vacuum as most gate parameters have small initial values. After 70 steps of optimization ($s=70$), the network learns to create a negative dip in the Wigner function that increases progressively after 120, 200, and 350 steps. The final stages of optimization refine the output state so that after 1000 steps it is almost indistinguishable from an ideal single photon with a fidelity of 99.998\%.}\label{Fig:Video}
\end{figure*}

In the following, we fix the non-Gaussian element $\Phi$ of each layer to be the Kerr gate $K=\exp(i\kappa \hat{n}^2)$, where $\hat{n}$ is the number operator. We choose the Kerr gate because it is diagonal in the Fock basis, a feature that leads to faster and more reliable numerical simulations. In the simulations, each mode is truncated to a subspace of fixed cutoff dimension, i.e., a state $\ket{\Psi}$ is represented in simulation by $\Pi_{D}\ket{\Psi}$, with $\Pi_{\mathcal{H}}$ the projector onto the $D$-dimensional truncated space. To prevent any initial, final, or intermediate state of the circuit having significant support outside of the space, we set the cutoff dimension to be suitably large to fully contain the state at all stages of the simulation, see Appendix~\ref{Appendix:CutoffVsRelations} for more details.

For simulations and experimental implementations, it is important to be aware of the magnitude of parameters in active gates that can change the photon number, namely the displacement, squeezing, and Kerr gates. It may be desirable to fix upper bounds on these parameters due to the simulation or hardware constraints. Tables \ref{tab:stateparameters} and \ref{tab:gateparameters} provide a summary of the various hyperparameters and results for each of the examples that follow in this paper, including in particular the largest parameters required for displacement, squeezing, and Kerr gates.

\section{State preparation}\label{Sec:StatePreparation}
The goal is to find a quantum circuit that, for a given canonical input state $\ket{\Psi_0}$ and a target output state $\ket{\Psi_t}$, performs a unitary transformation $U$ satisfying 
\beq\label{Eq: StatePrep}
\ket{\Psi_t}=U\ket{\Psi_0}.
\eeq
In other words, the circuit $U$ prepares the state $\ket{\Psi_t}$ from the input $\ket{\Psi_0}$. Eq.~\eqref{Eq: StatePrep} fixes a single column of the unitary $U$ in a basis containing the input state $\ket{\Psi_0}$, while the remaining components of $U$ are free parameters. For simplicity, we henceforth fix the input state to be the vacuum, i.e., $\ket{\Psi_0}=\ket{0}$.

In principle, it is possible to solve the state preparation problem by fixing the remaining free parameters of $U$ and finding an approximate decomposition in terms of a universal gate set. This approach has several drawbacks: it is not clear how to optimally choose the free parameters, finding decompositions can be computationally expensive, and the resulting circuits usually involve a very large number of gates from the universal set. These drawbacks are even more pronounced for CV quantum computers where finding decompositions is significantly more involved than in other models \cite{sefi2011decompose}.

Instead, our approach is inspired by strategies employed in machine learning. To prepare a given target state, we first fix the architecture by selecting the number of layers (depth) of the quantum neural network. Optimization is performed by minimizing the cost function
\beq\label{Eq: loss function}
C(\vec{\theta}) = \left|\bra{\Psi_t}U(\vec{\theta})\ket{0} - 1\right|,
\eeq
which captures the goal of finding parameters $\vec{\theta}$ such that $U(\vec{\theta})\ket{0}=\ket{\Psi_t}$, in which case $\bra{\Psi_t}U(\vec{\theta})\ket{0}=1$.

In the following, we showcase the effectiveness and versatility of this approach by selecting several canonical single and two-mode states and employing our techniques to find quantum circuits that prepare them. Our results demonstrate that the search for these circuits can be largely automated and the resulting circuits have short depth while still preparing the states with very high fidelity.

\subsection{Examples}

\subsubsection{Single photon state }

\begin{center}
\begin{figure}[t!]
\begin{tabular}{cc}
\includegraphics[width=0.5\columnwidth]{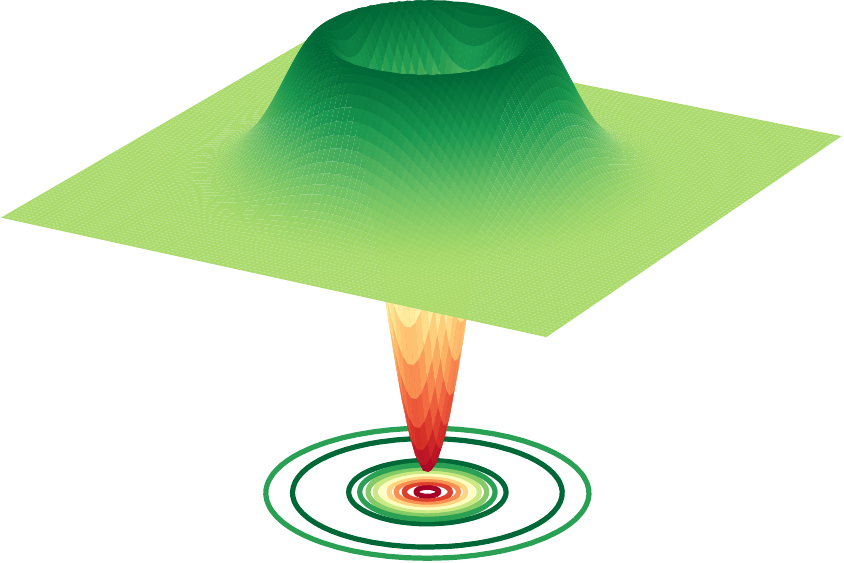}&
\includegraphics[width=0.5\columnwidth]{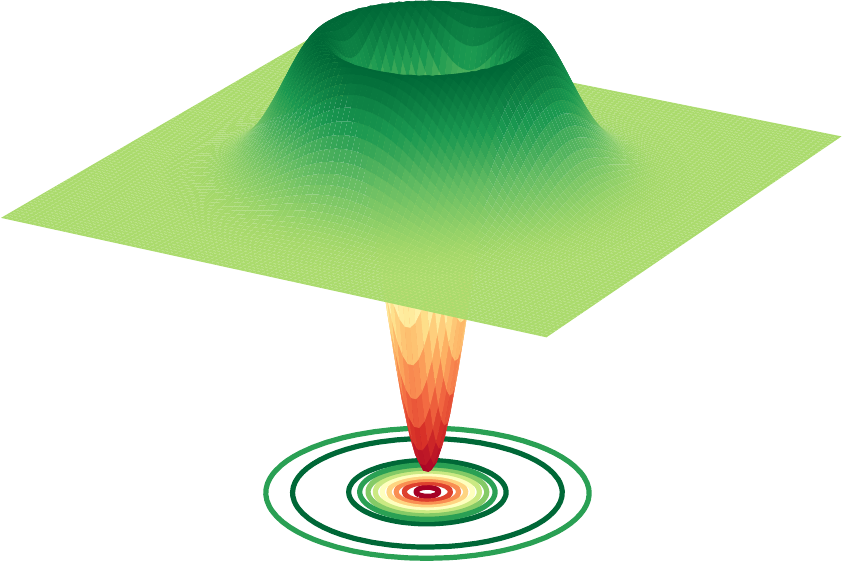}\\
\includegraphics[width=0.5\columnwidth, height=1.5cm]{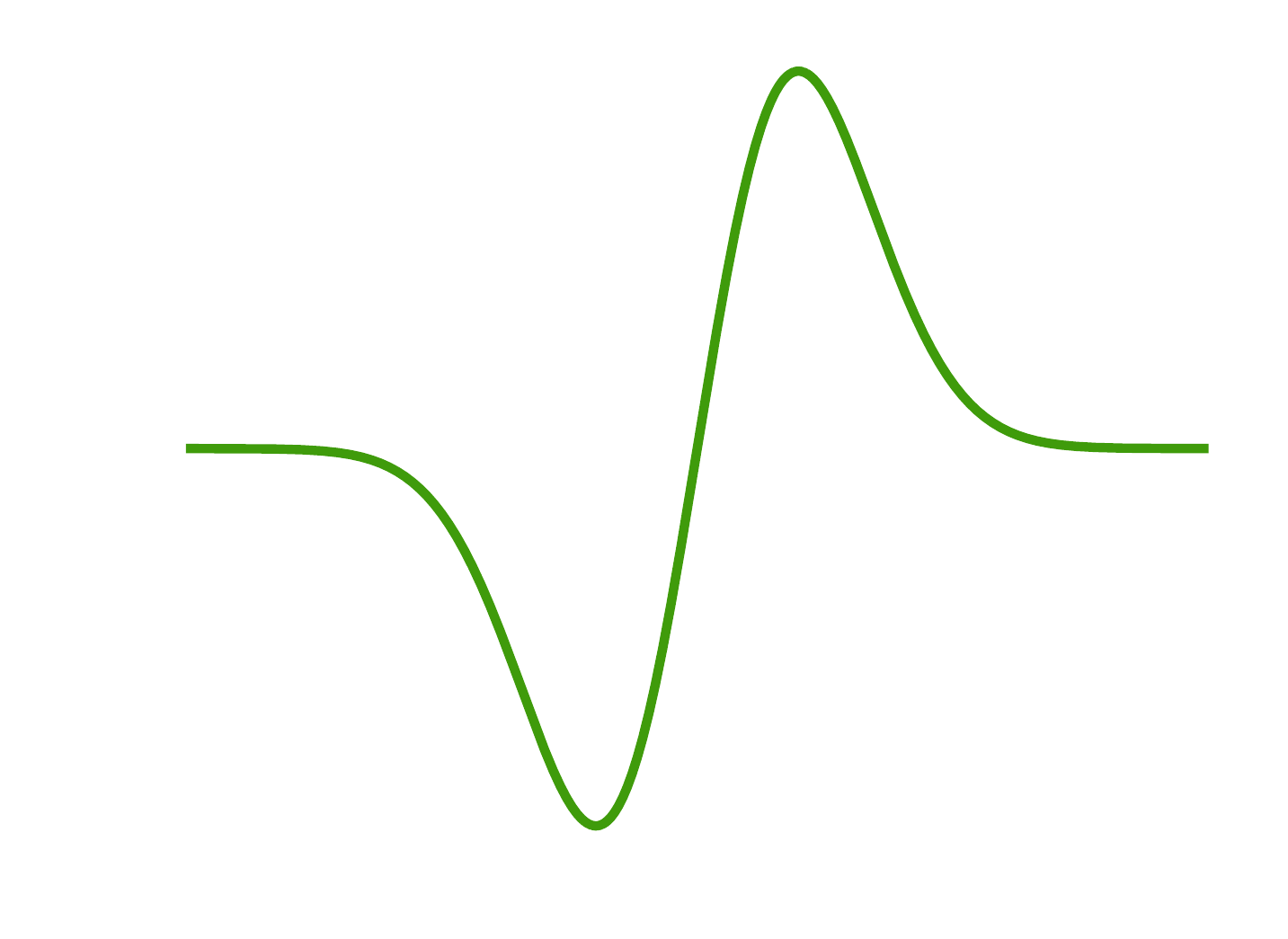}&
\includegraphics[width=0.5\columnwidth, height=1.5cm]{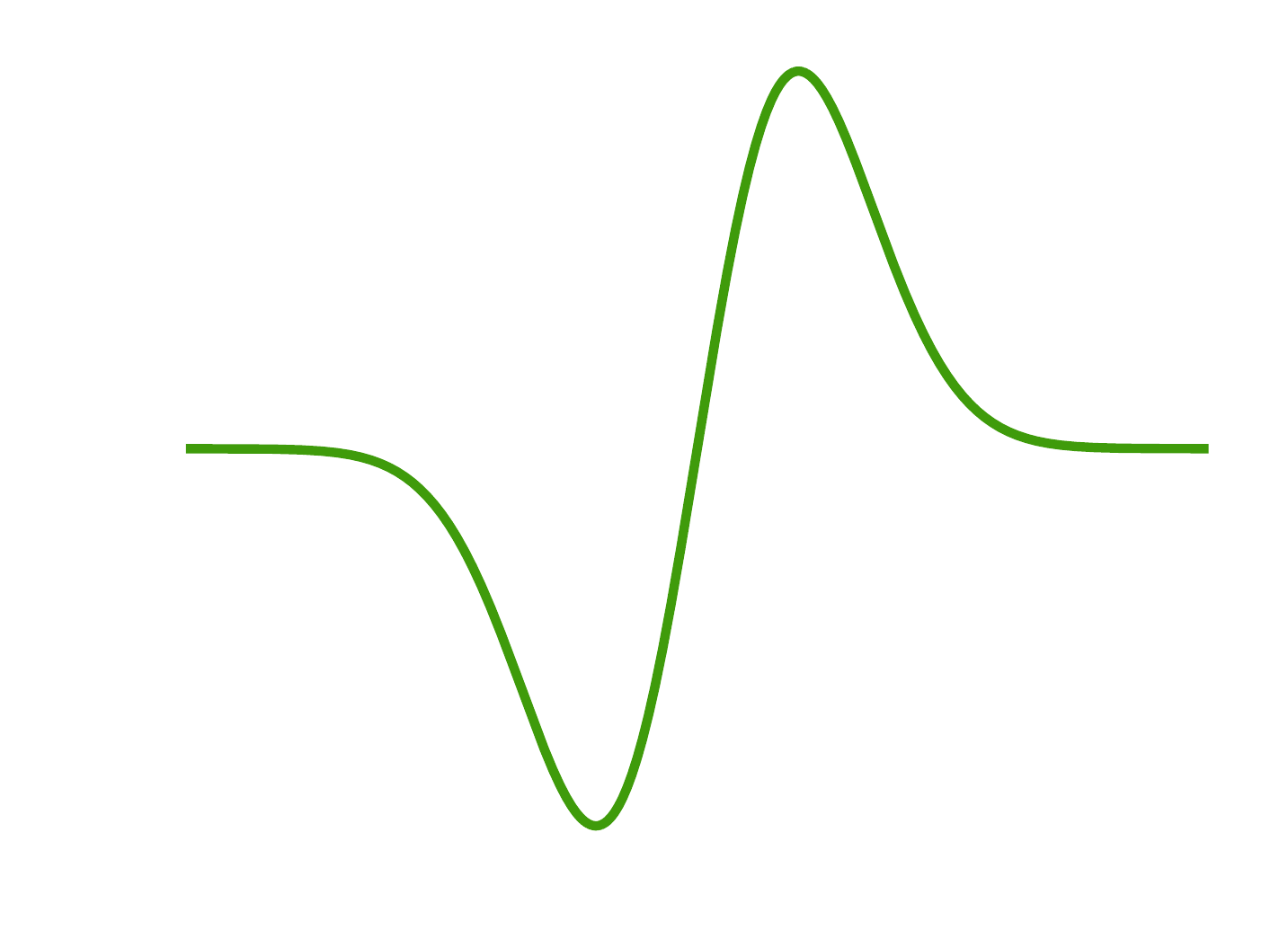}
\end{tabular}
\caption{(Top left) Wigner function of an ideal single photon. (Top right) Wigner function of the quantum state prepared by the quantum neural network. In both cases, a contour plot of the function is shown below. The fidelity between the ideal single photon and the prepared state is 99.998\%. (Bottom left) Wavefunction of the target. (Bottom right) Wavefunction of the prepared state. }\label{Fig:1photon}
\end{figure}
\end{center}

As a first example, we consider the task of preparing a single photon. We select a single-mode quantum neural network consisting of 8 layers, for a total of 40 gates from a universal set. For a single mode, a linear optical interferometer reduces to a rotation gate $R(\phi)=e^{i\phi\hat{n}}$. In Fig. \ref{Fig:Video}, we visualize the optimization process by displaying the Wigner function of the output state of the network at different stages during the optimization of the circuit. Initially, due to the random initialization of the circuit parameters, the network prepares a random state that is close to the vacuum. After only a few steps, the network learns to include negativity in the Wigner function. As the optimization continues, the network carves an output state that increasingly resembles an ideal single photon. The result of optimization is illustrated in Fig. \ref{Fig:1photon}, where we plot the Wigner functions of an ideal single photon and of the state prepared by the circuit. After 5000 steps of the gradient descent algorithm, a circuit is found that can prepare a state with 99.998\% fidelity to a perfect single photon. 

\begin{center}
\begin{figure}[t!]
\begin{tabular}{cc}
\includegraphics[width=0.85\columnwidth]{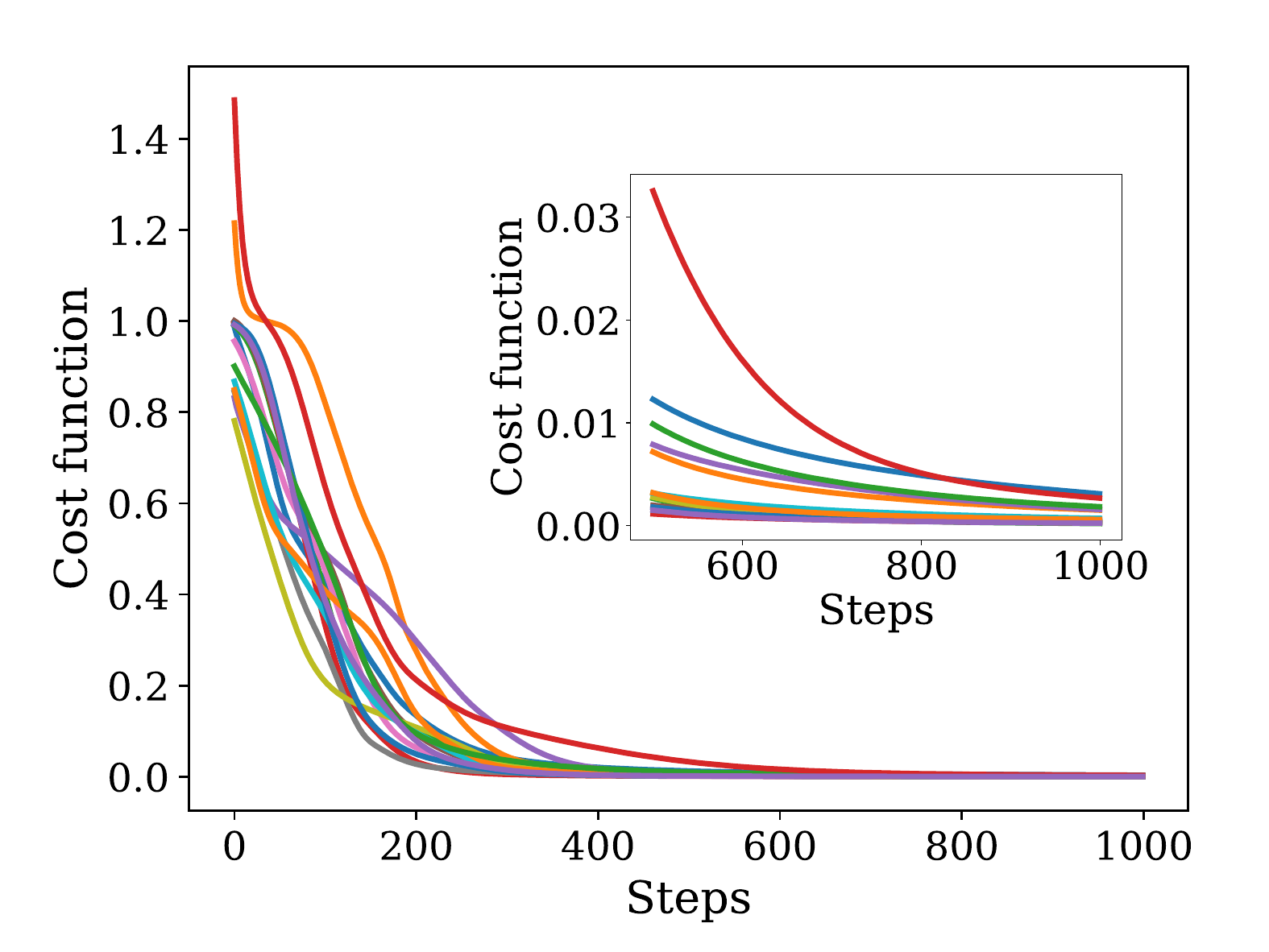}
\end{tabular}
\caption{Progress in minimizing the cost function for 15 independent runs of the optimization algorithm for preparing a single photon state. The inset shows the values of the cost function from 500 to 1000 steps. The randomness in the initialization of the circuit as well as in the stochastic gradient descent algorithm can result in different performances across each run. }\label{Fig:trajectories}
\end{figure}
\end{center}

\begin{center}
\begin{figure}[t!]
\begin{tabular}{cc}
\includegraphics[width=0.85\columnwidth]{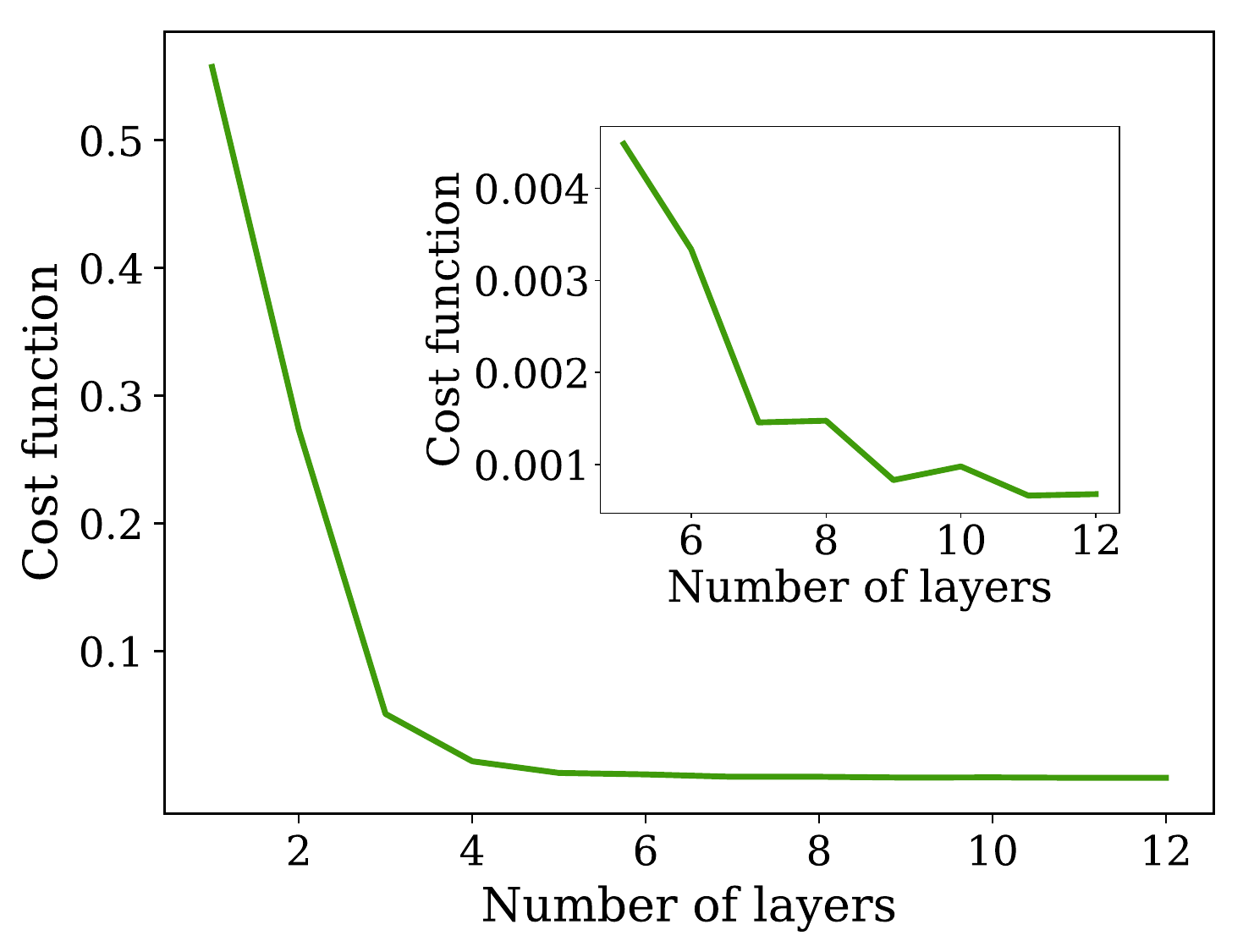}
\end{tabular}
\caption{Value of the cost function after 1000 steps of optimization for quantum neural networks with a different number of layers. Each value is an average over ten independent runs of the circuit optimization for preparing a single photon.}\label{Fig:layers}
\end{figure}
\end{center}

The Adam optimization algorithm is a stochastic gradient descent algorithm. Together with the random initialization of the gate parameters in the network, this leads to a non-deterministic output of the network. Although the optimization converges to low values of the cost function for a sufficiently large number of steps, for a fixed number of steps the values will differ across different independent runs, as shown in Fig. \ref{Fig:trajectories}. Consequently, it is often desirable to perform many independent optimization sessions and select the best output among them.

Finally, we remark that the depth of the quantum neural network plays an important role in the performance of state preparation. Complex quantum states require large circuits to prepare them and therefore a correspondingly large number of layers in the network. However, there is a price to pay for adding layers: it increases the number of parameters to optimize as well as the resources required to simulate the resulting circuits, leading to longer optimization times. It is therefore preferable to find an adequate number of layers such that sufficiently high fidelities can be reached while minimizing the computational resources required for optimization. This is shown in Fig. \ref{Fig:layers}, where we find that the single photon state can be well approximated using only a few layers. As demonstrated also in subsequent examples for both state preparation and gate synthesis, it is possible to find circuits of remarkably short depth that can still achieve high fidelities.

\subsubsection{ON state}  
\begin{center}
\begin{figure}[t!]
\begin{tabular}{cc}
\includegraphics[width=0.5\columnwidth]{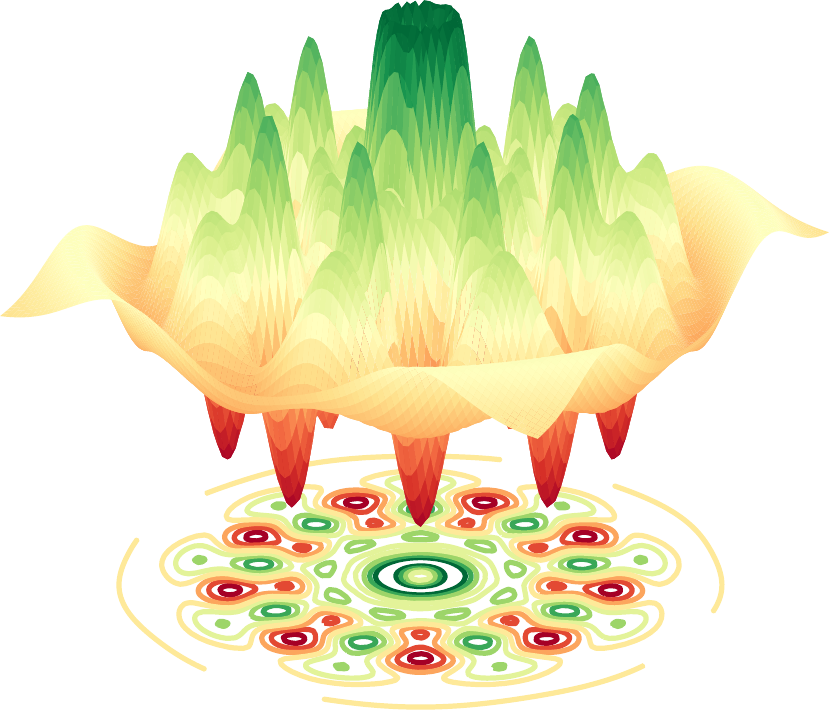}&
\includegraphics[width=0.5\columnwidth]{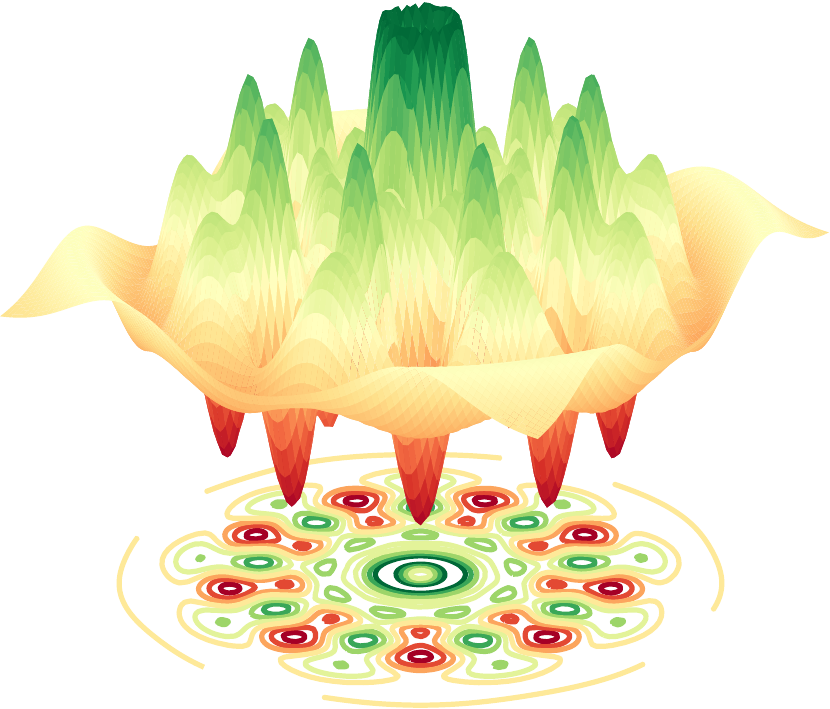}\\
\includegraphics[width=0.5\columnwidth, height=1.5cm]{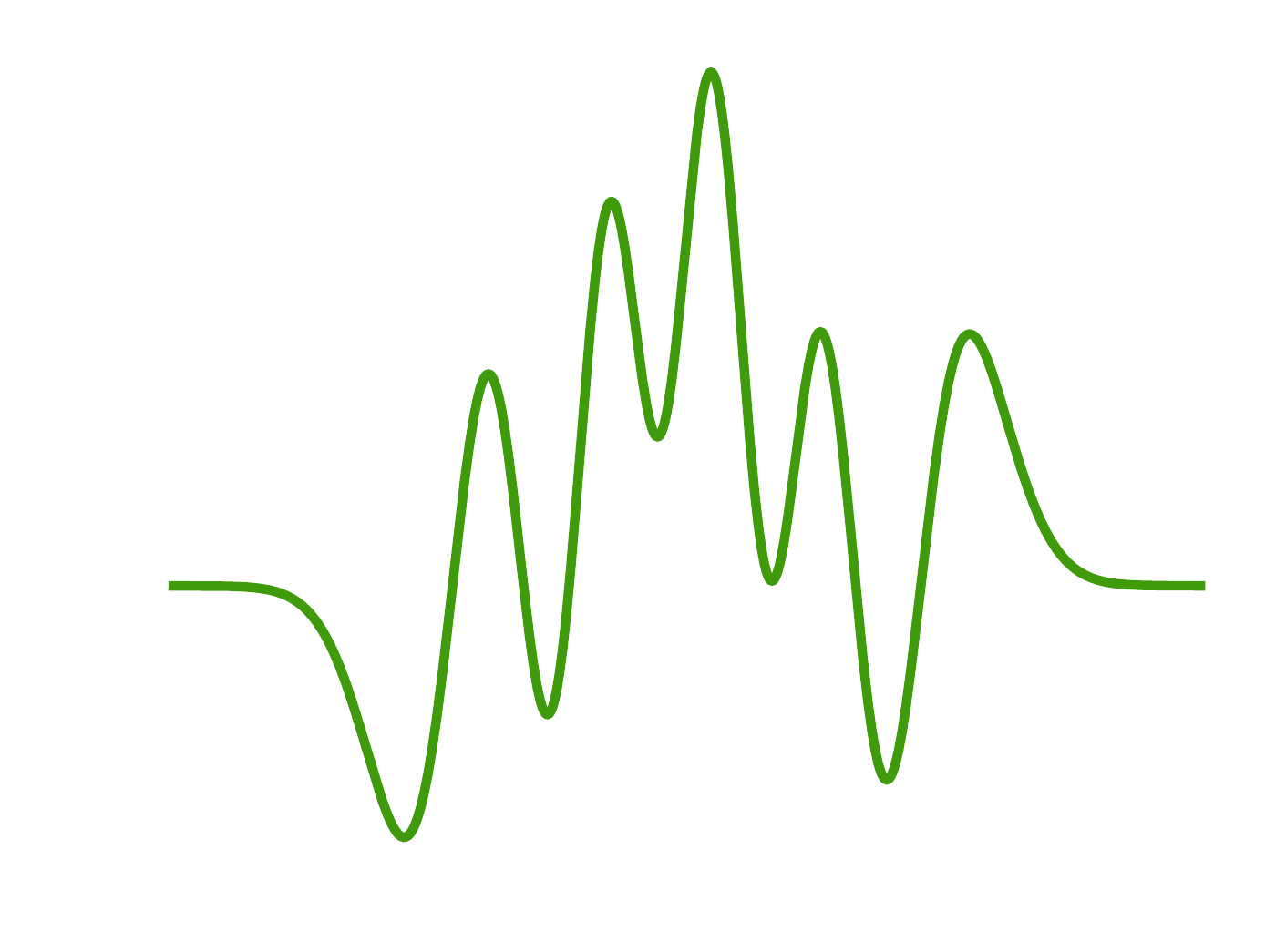}&
\includegraphics[width=0.5\columnwidth, height=1.5cm]{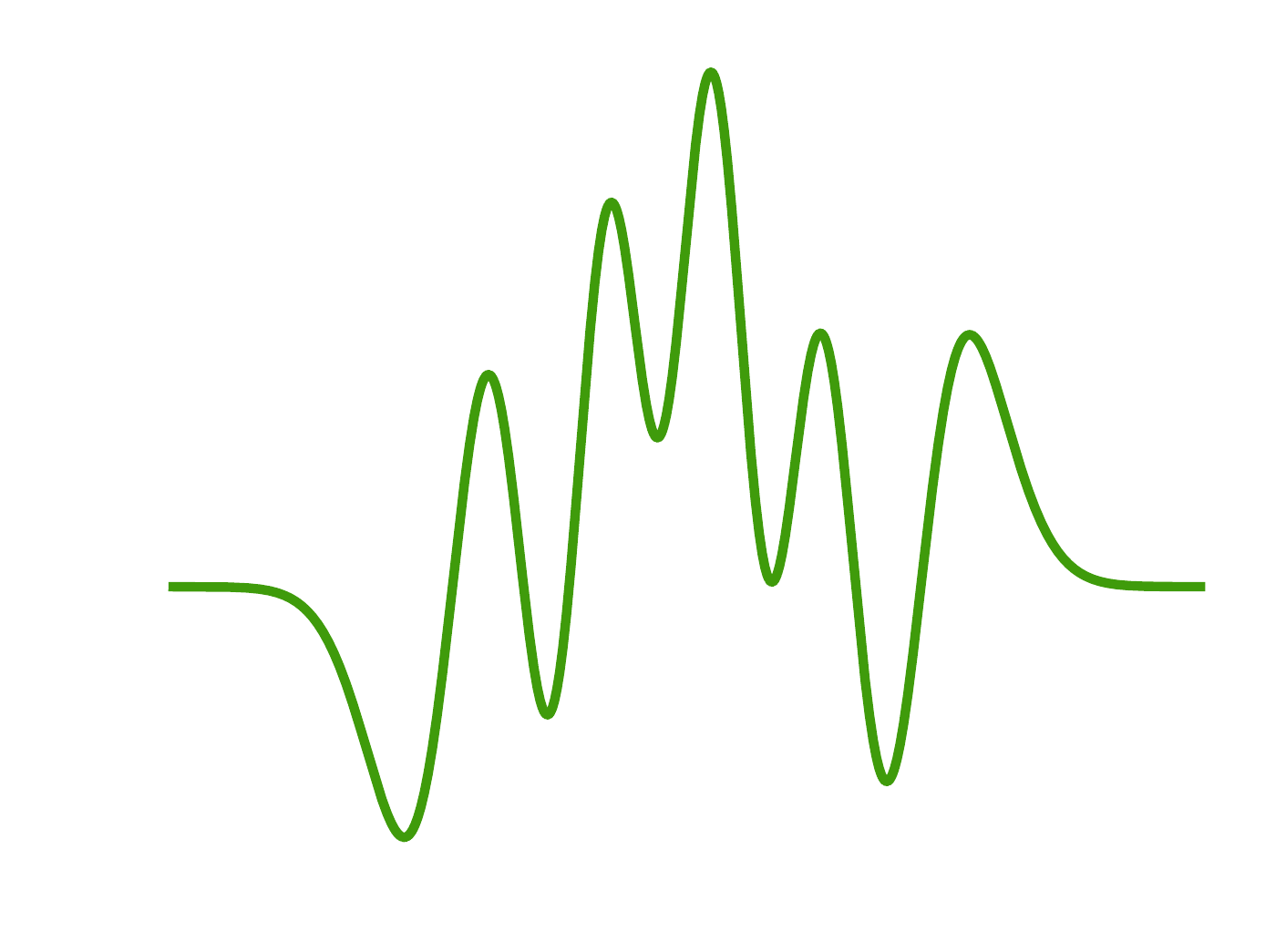}
\end{tabular}
\caption{ (Top left) Wigner function of an ideal ON state with $N=9$. (Top right) Wigner function of the quantum state prepared by the quantum neural network. In both cases, a contour plot of the function is shown at the bottom of the figure. The fidelity between the ideal ON state and the prepared state is 99.93\%. (Bottom left) Wavefunction of the target state. (Bottom right) Wavefunction of the prepared state.}\label{Fig:ON}
\end{figure}
\end{center}

In the measurement-based model of CV quantum computing \cite{menicucci2006universal,gu2009quantum}, quantum gates are applied via gate teleportation: a technique where measuring a resource state in a neighbouring mode enables the application of a corresponding gate on a target mode. In Ref. \cite{sabapathy2018states}, it was shown that superpositions of vacuum and Fock states of the form 
\beq
\ket{\Psi_{\text{ON}}}=\frac{1}{\sqrt{1+|a|^2}}(\ket{0}+a\ket{N}),
\eeq
where $\ket{0}$ is the vacuum, $\ket{N}$ is an $N$-photon Fock state, and $a$ is a complex number, can be employed via gate teleportation to apply the gate $\exp(i\tau \hat{x}^N)$ to first order in $\tau$. These states are known as ON states and we select the $N=9$ ON state as the target for state preparation, setting $a=1$ such that the state is an equal superposition of the vacuum and a nine-photon Fock state. We fix a network of 20 layers (100 gates) and optimize for 5000 steps, using a cutoff dimension of 14. The result is a circuit that can prepare a state with 99.93\% fidelity to the ideal ON state as shown in Fig. \ref{Fig:ON}.

\subsubsection{GKP state} Gottesman-Kitaev-Preskill (GKP) states \cite{gottesman2001encoding} are important resource states that can be used for error correction in CV quantum computing \cite{menicucci2014fault}. Here, we focus on Hex GKP states, as defined in Ref. \cite{noh2018improved}, which are better suited than the original GKP states for correcting errors due to loss. In their idealized form, Hex GKP states are given by
\begin{align}
|\mu\rangle \propto  \sum_{n_{1},n_{2}=-\infty}^{\infty}  e^{ -i(\frac{1}{2}\hat{q}+\frac{\sqrt{3}}{2}\hat{p})\sqrt{\frac{4\pi}{\sqrt{3}d}} (dn_{1}+\mu) }   e^{ i\hat{q}\sqrt{\frac{4\pi}{\sqrt{3}d}} n_{2} } |0\rangle,
\label{eq:52} 
\end{align}
where $d$ is the dimension of a code space, $\mu=0,1,\ldots, d-1$, and $\ket{0}$ is the vacuum. In the case of finite energy, the states are modulated by a Gaussian envelope $\exp(-\Delta^{2}\hat{n})\ket{\mu}$.

\begin{figure}[t!]
\begin{tabular}{cc}
\includegraphics[width=0.5\columnwidth]{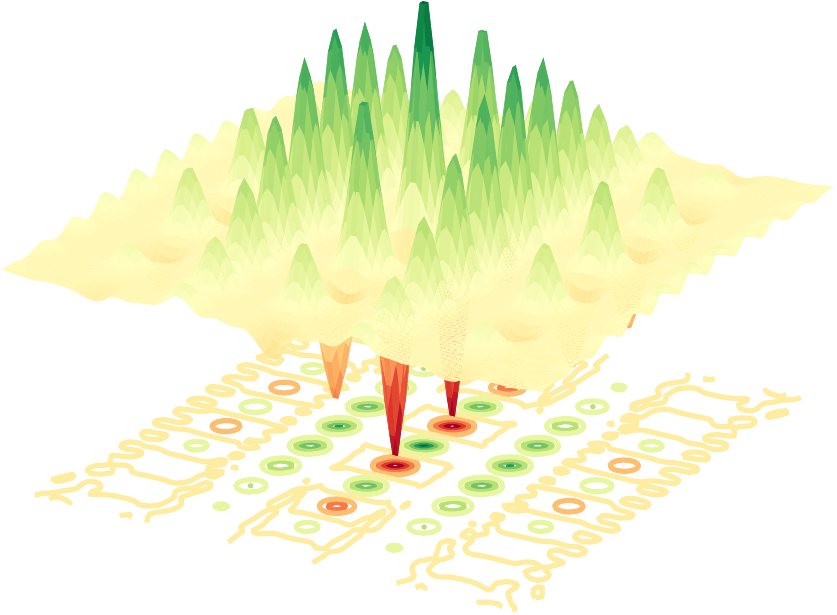}&
\includegraphics[width=0.5\columnwidth]{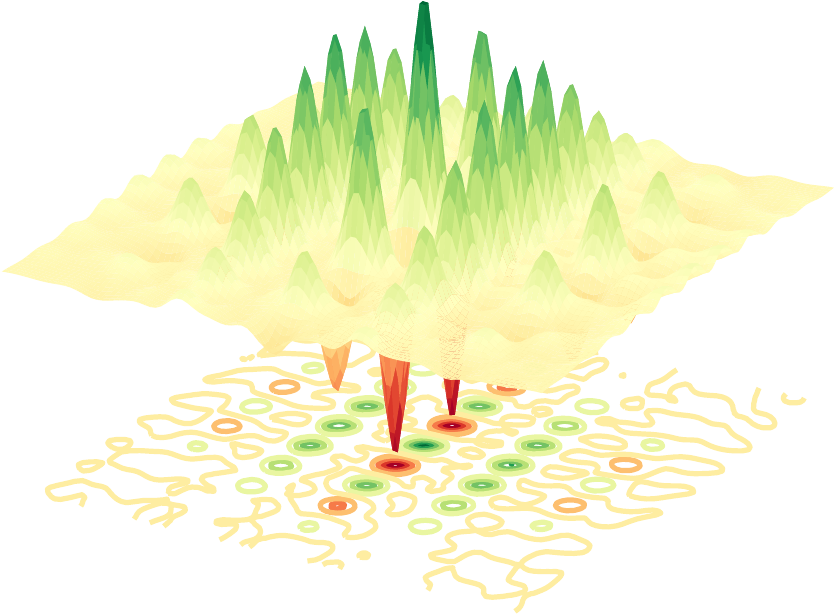}\\
\includegraphics[width=0.5\columnwidth, height=1.5cm]{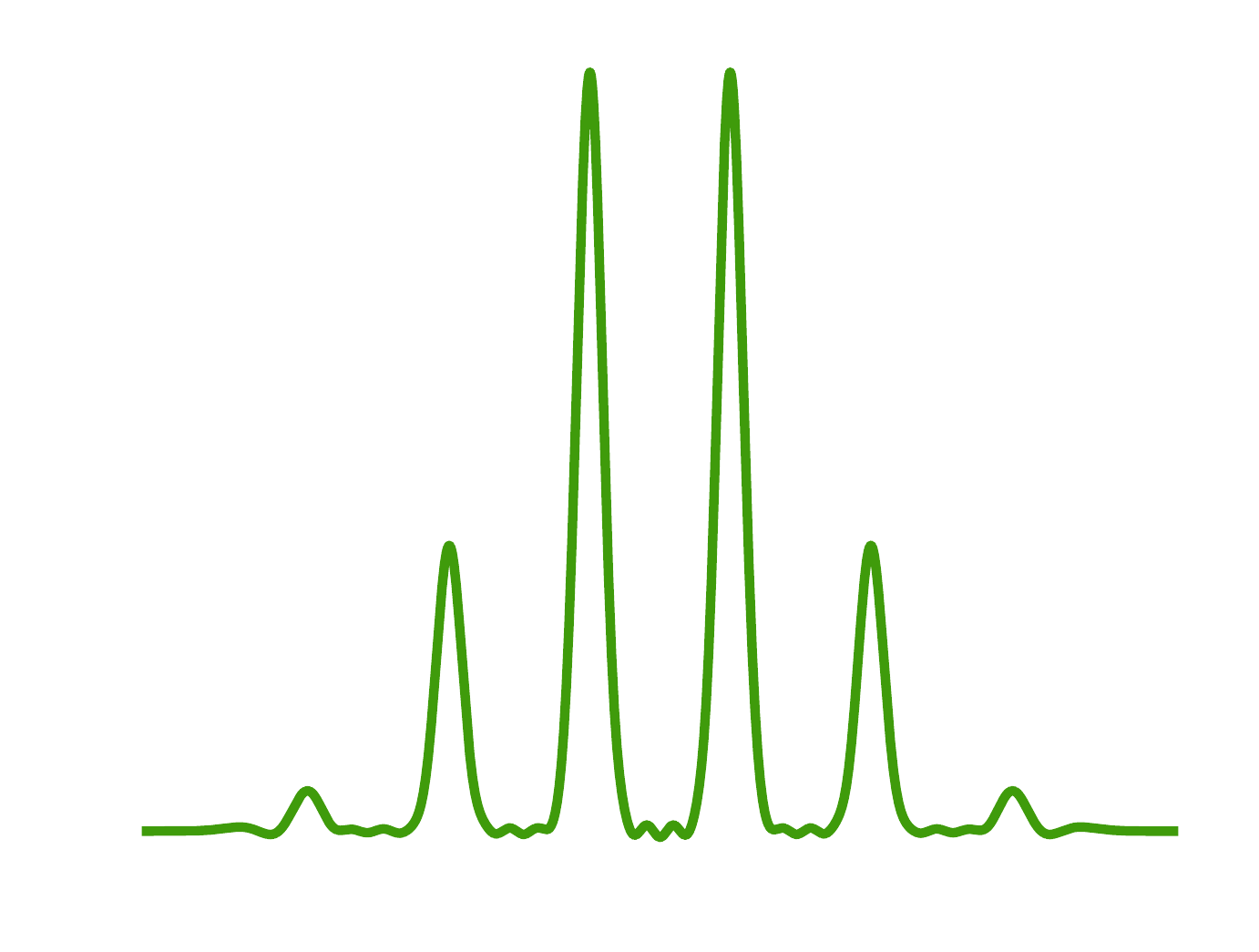}&
\includegraphics[width=0.5\columnwidth, height=1.5cm]{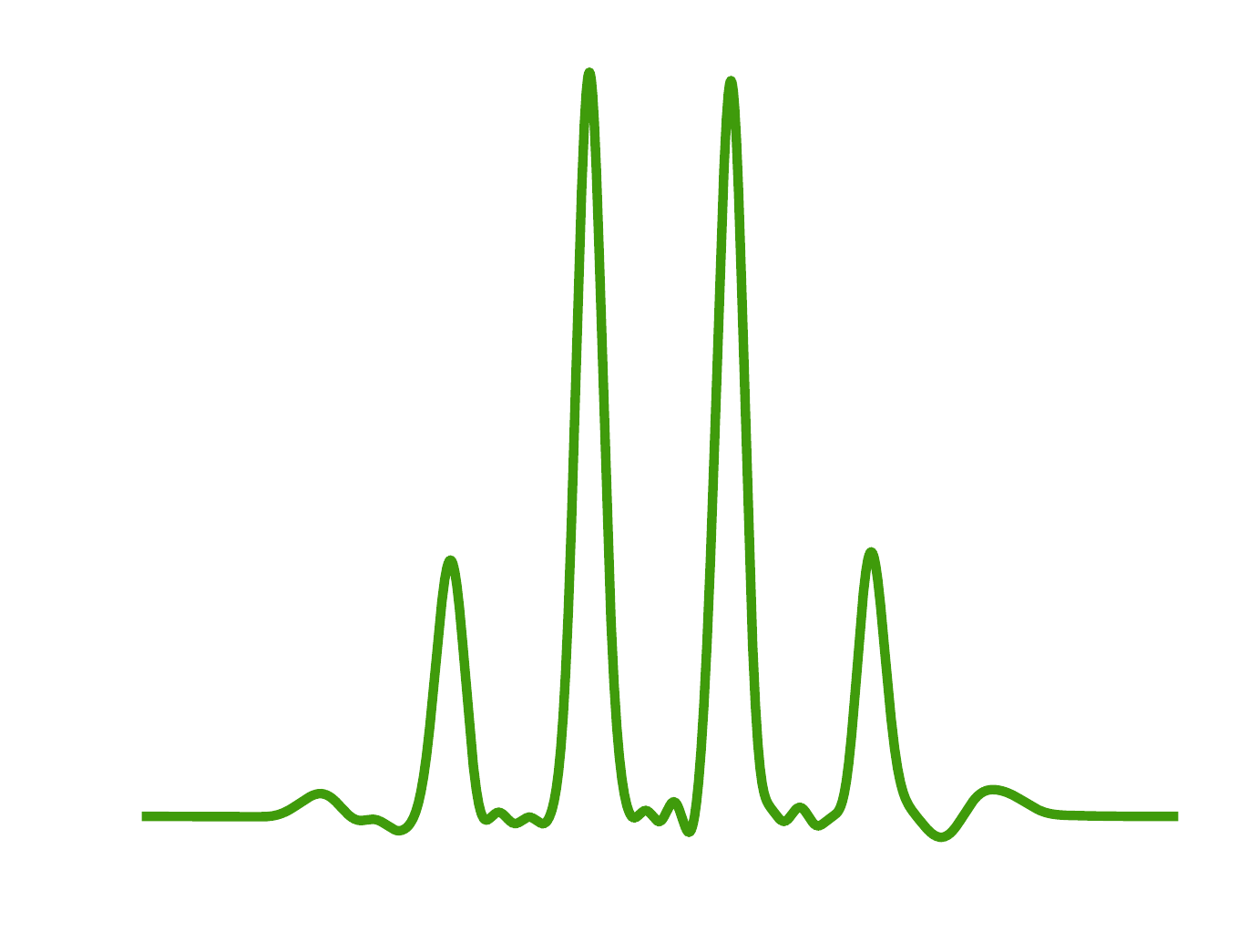}
\end{tabular}
\caption{(Top left) Wigner function of an ideal Hex GKP state with parameters $\mu=1$ and $\Delta=0.3$. (Top right) Wigner function of the quantum state prepared by the quantum neural network. In both cases, a contour plot of the function is shown at the bottom of the figure. The fidelity between the ideal Hex GKP state and the prepared state is 99.60\%. (Bottom left) Wavefunction of the target state. (Bottom right) Wavefunction of the prepared state.}\label{Fig:GKP} 
\end{figure}

In Strawberry Fields, it is convenient to specify states in terms of their components in the Fock basis. In the case of the GKP states, these can be computed by numerically evaluating the inner products $\braket{\Psi}{n}$ for all Fock states $\ket{n}$ in the Hilbert space defined by the cutoff dimension in the simulation. We choose a Hex GKP state with $\mu=1$, and $\Delta=0.3$ as the target for state preparation. This state is significantly more complex than our previous examples, so we select a network with 25 layers (125 gates), a cutoff dimension of 51 for the simulation, and optimize for 10,000 steps. The result is illustrated in Fig. \ref{Fig:GKP}, where a circuit was found that can prepare a state with 99.60\% fidelity to the target GKP state.

\subsubsection{Random state} 
\begin{center}
\begin{figure}[t!]
\begin{tabular}{cc}
\includegraphics[width=0.5\columnwidth]{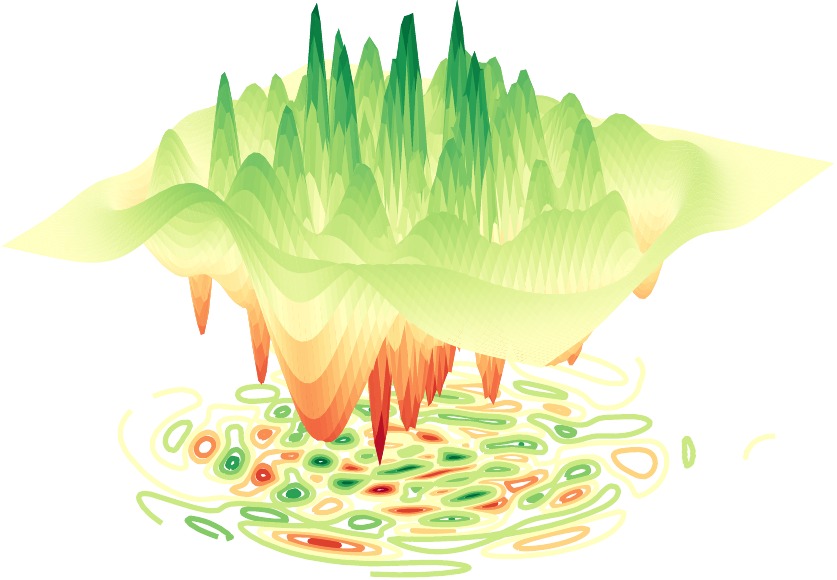}&
\includegraphics[width=0.5\columnwidth]{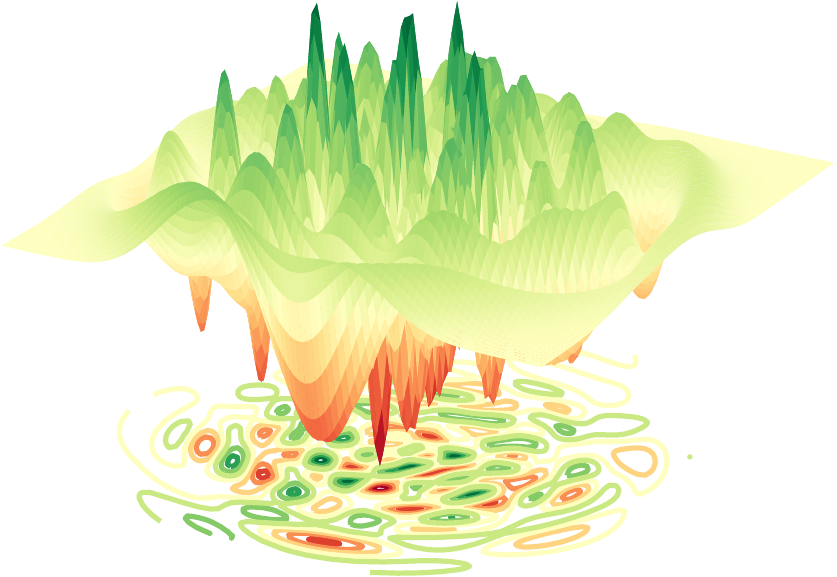}\\
\includegraphics[width=0.5\columnwidth, height=1.5cm]{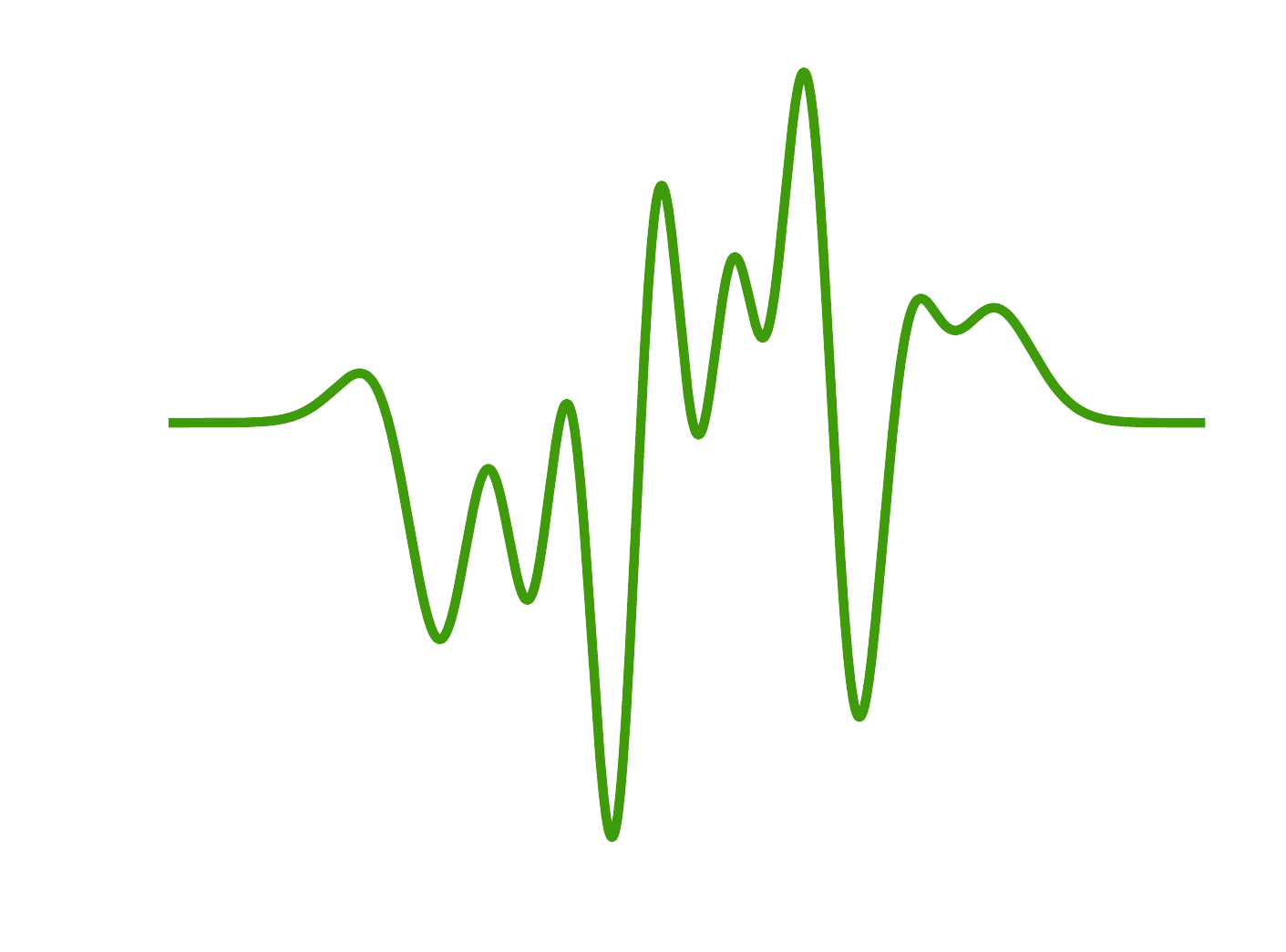}&
\includegraphics[width=0.5\columnwidth, height=1.5cm]{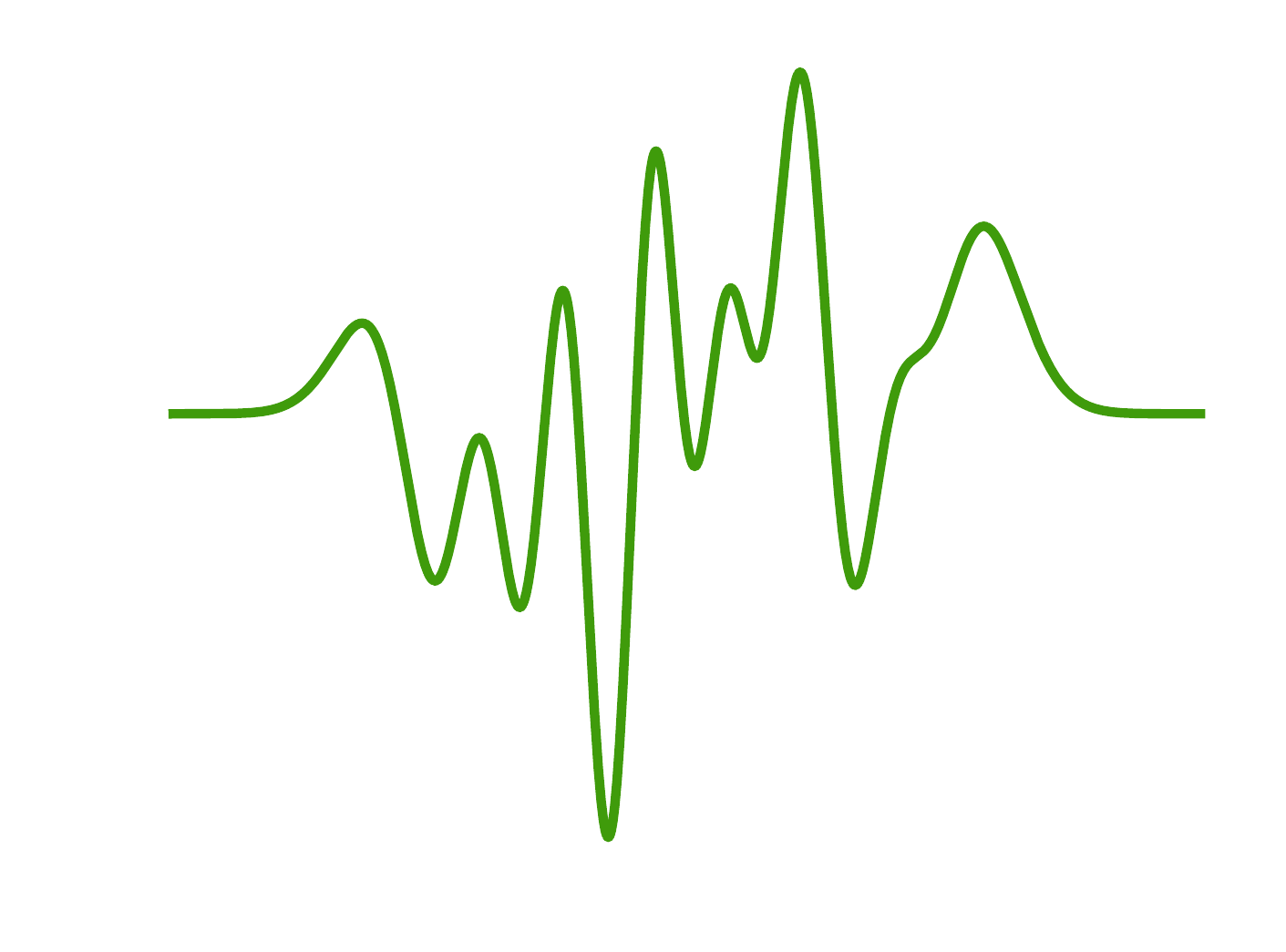}
\end{tabular}
\caption{(Top left) Wigner function of the target random state of Eq. \eqref{Eq: randstate} with $d=15$. (Top right) Wigner function of the quantum state prepared by the quantum neural network. In both cases, a contour plot of the function is shown at the bottom of the figure. The fidelity between the target state and the prepared state is 99.82\%. (Bottom left) Wavefunction of the target state. (Bottom right) Wavefunction of the prepared state.}\label{Fig:randomState} 
\end{figure}
\end{center}

All the states studied thus far have significant symmetry. To demonstrate that our method is versatile enough to prepare unstructured states, we now focus on a random state of the form
\beq\label{Eq: randstate}
\ket{\Psi_R}=\frac{1}{\mathcal{N}}\sum_{n=0}^{d-1}(a_n+ib_n)\ket{n},
\eeq 
where the values $a_n$ and $b_n$ where drawn from a standard normal distribution and $\mathcal{N}=\sum_{n=0}^{d\tb{-1}}|(a_n+ib_n)|^2$ is a normalization constant. We fix a value of $d=15$ and select a network of 25 layers (125 gates), with a cutoff dimension of 20 in the simulation. The optimization is carried out for 5000 steps resulting in a prepared state with fidelity 99.82\% with respect to the target random state. This is shown in Fig. \ref{Fig:randomState}. 

\subsubsection{NOON state} As a final example, we extend the state preparation method to two-mode states. Computationally, optimization is significantly more challenging because the Hilbert space dimension in the simulation is quadratically larger than in the single-mode case. As the target state we select a NOON state, defined as
\beq
\ket{\Psi_\text{NOON}}=\frac{1}{\sqrt{2}}(\ket{N0}+\ket{0N}),
\eeq
where $\ket{0N}$ is a state of vacuum in the first mode and $N$ photons in the second mode, with an analogous definition for the state $\ket{N0}$. NOON states play an important role in quantum metrology as they are known to be optimal resource states that saturate the Heisenberg limit of precision for phase estimation \cite{sanders1989quantum, boto2000quantum, giovannetti2006quantum}.

\begin{center}
\begin{figure}
\begin{tabular}{cc}
\includegraphics[width=0.5\columnwidth]{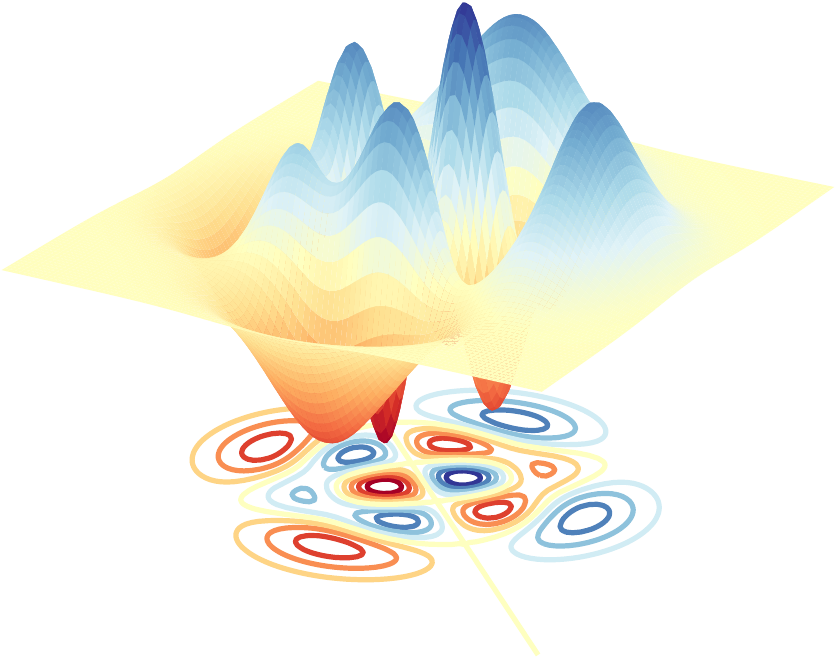}&
\includegraphics[width=0.5\columnwidth]{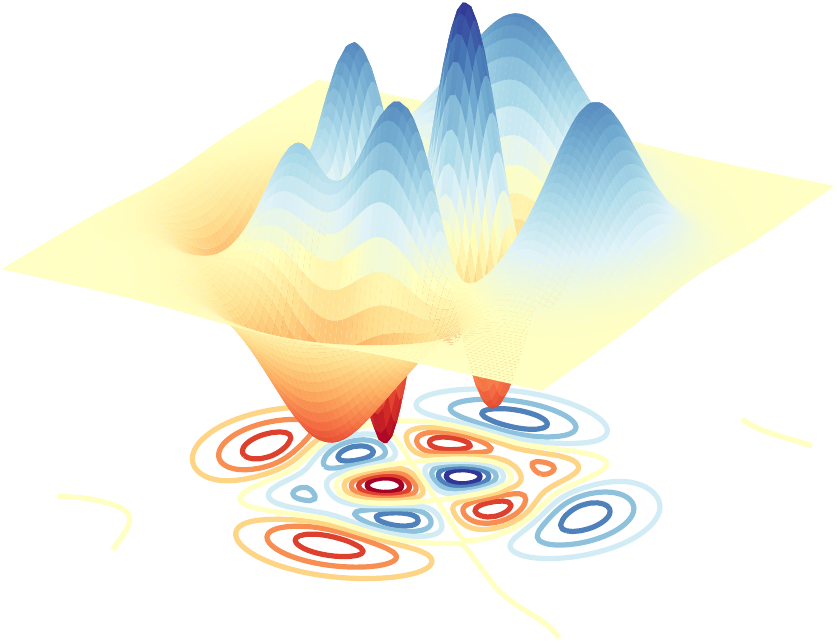}
\end{tabular}
\caption{(Left) Two-dimensional wavefunction $\Psi_{\text{NOON}}(x,y)$ of a NOON state with $N=5$. (Right) Two-dimensional wavefunction of the quantum state prepared by the quantum neural network. The fidelity between the target state and the prepared state is 99.9\%. }\label{Fig:NOON} 
\end{figure}
\end{center}

We set the NOON state with $N=5$ as the target state, fixing a network of 20 layers (100 gates) with a cutoff dimension of 10. The network prepares a state with 99.89\% fidelity to the ideal target state. This is shown in Fig. \ref{Fig:NOON}, where we plot the two-dimensional wavefunction $\Psi_{\text{NOON}}(x,y)$ of the target and prepared states. Note that we plot the wavefunction and not the Wigner function since the latter is a function of four variables in the two-mode case. 

\begin{table*}[t!]
	\centering
	\begin{ruledtabular}
		\begin{tabular}{lccccc}
		
			\textit{Hyperparameters} &\textit{ Single photon state} & \textit{ON state} & \textit{Hex GKP state} & \textit{Random state} &\textit{ NOON state}\\\hline
			Fock basis cutoff dimension $D$ & 6 & 14 & 50 & 20 &10\\
			Depth/number of layers & 8 & 20 & 25 & 25 & 20\\
			Total elementary gate count & 40 & 100 & 125 & 125 & 100\\
			Number of optimization steps & 5000 & 5000 & 10000 & 5000 & 5000\\
			Target state parameters & $-$ & $a=1,N=9$ & $\mu=1,\Delta=0.3$ & $-$ & $N=5$ \\
			\hline
			\textit{Results and circuit parameters} & & & &\\\hline
			Fidelity to the target state $|\braket{\Psi_t}{\Psi}|^2$ & $99.998\%$ & $99.93\%$ & $99.60\%$ & $99.82\%$ & $99.89\%$\\
			Runtime (seconds) & 65 & 436 & 6,668 & 1,061 & 1,270\\
			Maximum absolute displacement $|\alpha|$ & $0.4481$ & $0.4136$ &$0.2590$ & $0.2188$ &$0.5036$\\
			Maximum absolute squeezing $|r|$ & $0.09728$ & $0.2355$ &$0.2793$ & $0.2143$ &$0.2533$\\
			Maximum absolute Kerr strength $|\kappa|$ & $0.5132$ & $0.4110$ & $0.2484$ & $0.2654$ & $0.4129$
		\end{tabular}
	\end{ruledtabular}
Optimization was run on a quadcore AMD Radeon R7 computer operating at 2.1GHz with 12GB RAM for all cases except the Hex GKP state, where the optimization was performed on a 20-core Intel Xeon CPU operating at 2.4GHz with 252GB RAM.
	\caption{Summary of state preparation hyperparameters and learning results. }
	\label{tab:stateparameters}
\end{table*}

All state preparation results are summarized in Table~\ref{tab:stateparameters}. A straightforward generalization of state preparation is gate synthesis, where we consider the action of the circuit not just on a fixed input, but on a complete orthonormal basis of a given Hilbert space. In the following section, we describe how our methods can be extended to the gate synthesis problem. 

\section{Gate synthesis}
\label{sec:gate}

In the context of photonic quantum computing, a gate is a quantum transformation operating either on the full infinite-dimensional Hilbert space or on a restricted subspace of a set of modes. Gates are most often of unitary form, which is the setting we consider in this work. One approach to synthesize a target unitary gate $V_t$ is to decompose it into a fixed set of elementary gates. The set of gates in the elementary set can vary, and there exists multiple decomposition methods for each gate set~\cite{lloyd1999quantum,sefi2011decompose}. However, there are several drawbacks to traditional methods for performing gate decomposition, including computational overhead, a large number of gates from the elementary set, and errors arising due to Trotterization and commutator approximations.

As with state preparation, we approximate $V_t$ using a CV quantum neural network whose parameters are optimized to replicate the action of the target gate. This process allows for a realization of the target unitary $V_t$ that is both deterministic and of fixed depth, providing the user with control over the number of elementary gates used. The strategy is as follows. Any given unitary can be equivalently described by a set of input-output relations between basis states: given a set of orthogonal input states $\{|\Psi_{0}^{(0)}\rangle,\ldots,|\Psi_{0}^{(d)}\rangle\}$, the action of $V_t$ can be described by specifying a set of target states $|\Psi_{t}^{(i)}\rangle = V|\Psi_{0}^{(i)}\rangle$ for all $i=0,\ldots,d-1$.

We fix the input states to be the first $d$ states in the Fock basis of each mode, i.e., $|\Psi_{0}^{(i)}\rangle=\ket{i}$ where $\ket{i}$ is the Fock state of $i$ photons. For unitaries acting on the infinite-dimensional multimode space, this represents a restriction on the countably infinite number of relations. This description of the unitary captures its action only on a subspace of restricted photon number. Energy constraints in any physical implementation of a CV quantum computer mean that an effective restriction on photon number is already introduced, and the user can fix $d$ to match this limitation. 

To optimize the unitary transformation $U(\vec{\theta})$ performed by the quantum neural network, we opt for the cost function

\begin{align}
C(\vec{\theta})&=\frac{1}{d}\sum_{i=0}^{d-1}\left|\langle\Psi_t^{(i)}|U(\vec{\theta})\ket{i} - 1\right|,\nonumber\\
&=\frac{1}{d}\sum_{i=0}^{d-1}\left|\bra{i}V_t^{\dagger}U(\vec{\theta})\ket{i} - 1\right|.
\end{align}
By comparing to the cost function in Eq.~\eqref{Eq: loss function}, we see that gate synthesis can be considered as a generalization of state preparation to more than one input-output relation.

The following section provides example applications of gate synthesis to several single and two-mode gates. These results demonstrate how relatively short-depth approximations of a unitary can be constructed using a process that is largely automated.

We measure the performance of gate synthesis by numerically calculating the average fidelity between the target unitary $V_t$ and the circuit unitary $U(\vec{\theta})$ when applied over all states with support in the input $d$-dimensional Fock subspace; see Appendix~\ref{Appendix:AvgMinFidelity} for more details. 

\subsection{Examples}

\subsubsection{Cubic phase gate}

The cubic phase gate is a single mode non-Gaussian unitary that, in combination with Gaussian gates, forms a universal set for CV quantum computing~\cite{weedbrook2012gaussian}. It is defined as
\beq
V_{\text{CPG}}(\gamma)=\exp \left(- i \gamma \hat{x}^{3}\right),
\eeq
with $\hat{x}$ the position operator and $\gamma>0$ a fixed gate parameter. As discussed previously, we synthesize a gate that replicates the action of the cubic phase gate on a subspace of the infinite-dimensional Hilbert space. We set this subspace to be the 10-dimensional subspace of states with at most nine photons and fix a gate parameter of $\gamma=0.01$. We select a network consisting of 25 layers, i.e., 125 gates, and perform the optimization for $4000$ steps.

\begin{center}
\begin{figure}[t!]
\begin{minipage}{0.22\columnwidth}
\includegraphics[width=\columnwidth]{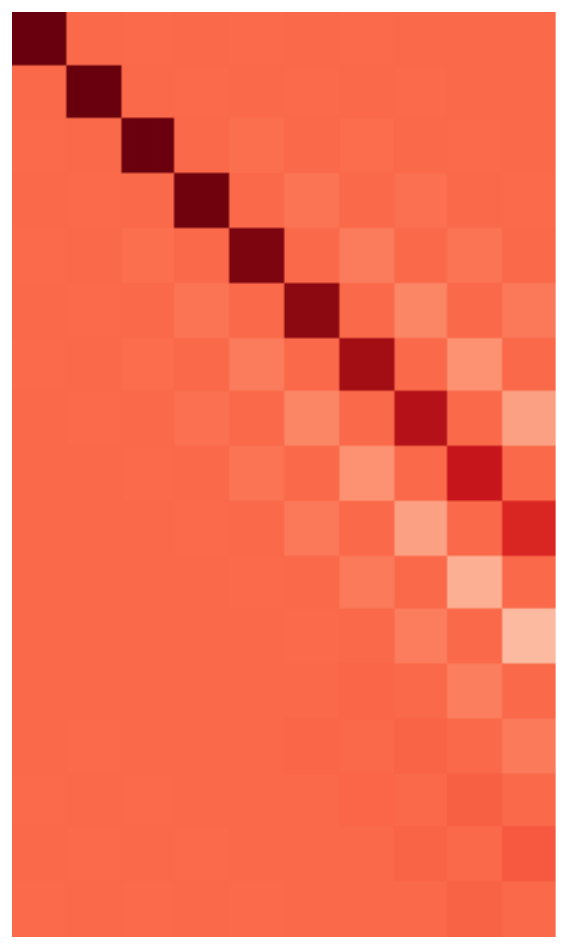}
\end{minipage}
\hspace{-7pt}
\begin{minipage}{0.22\columnwidth}
\includegraphics[width=\columnwidth]{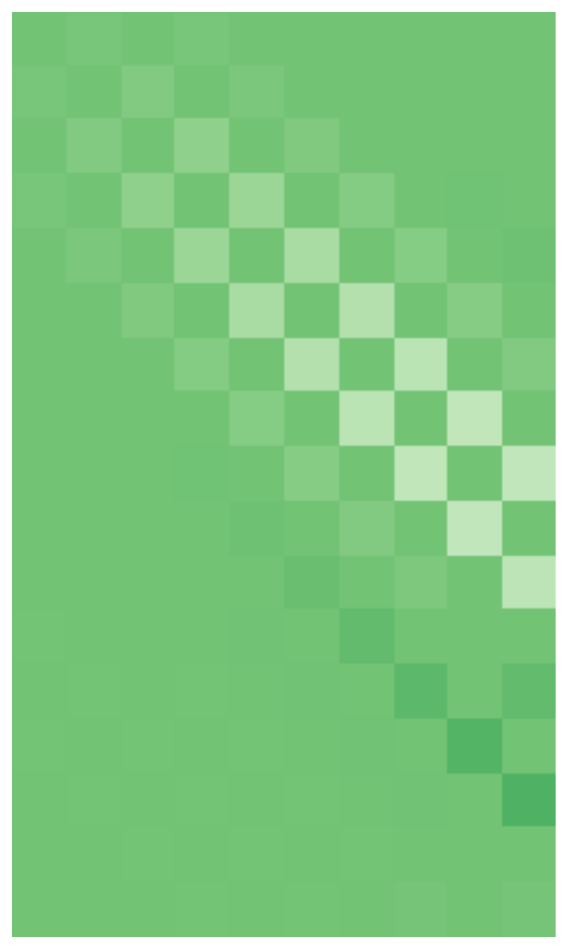}
\end{minipage}
\hspace{5pt}
\begin{minipage}{0.22\columnwidth}
\includegraphics[width=\columnwidth]{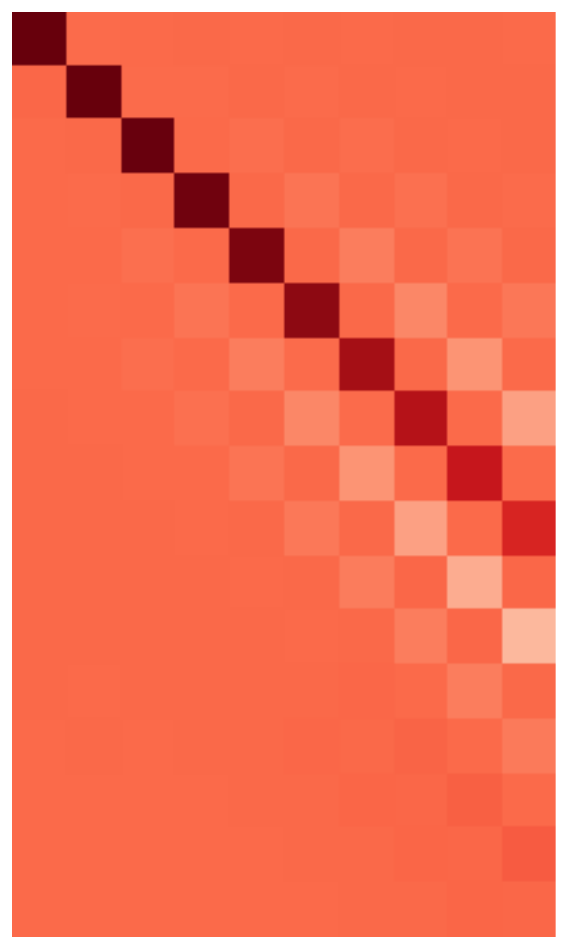}
\end{minipage}
\hspace{-7pt}
\begin{minipage}{0.22\columnwidth}
\includegraphics[width=\columnwidth]{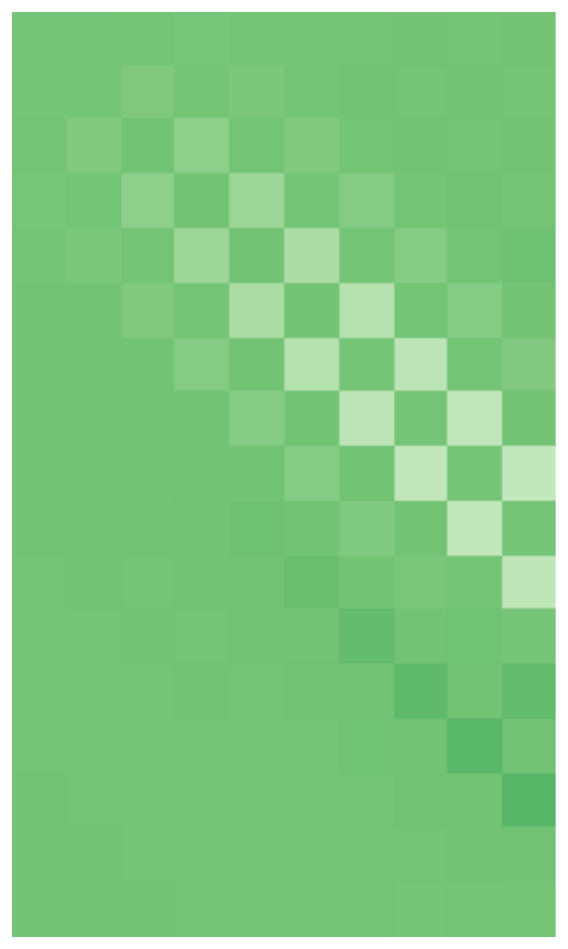}
\end{minipage}
\begin{minipage}{0.98\columnwidth}
\vspace{0.5cm}
\begin{tabular}{cc}
\includegraphics[width=0.48\columnwidth]{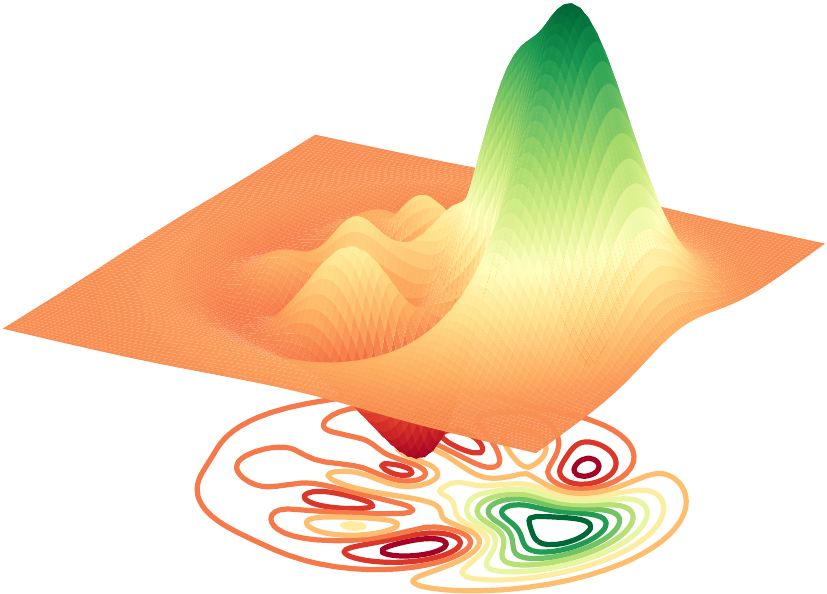}&
\includegraphics[width=0.48\columnwidth]{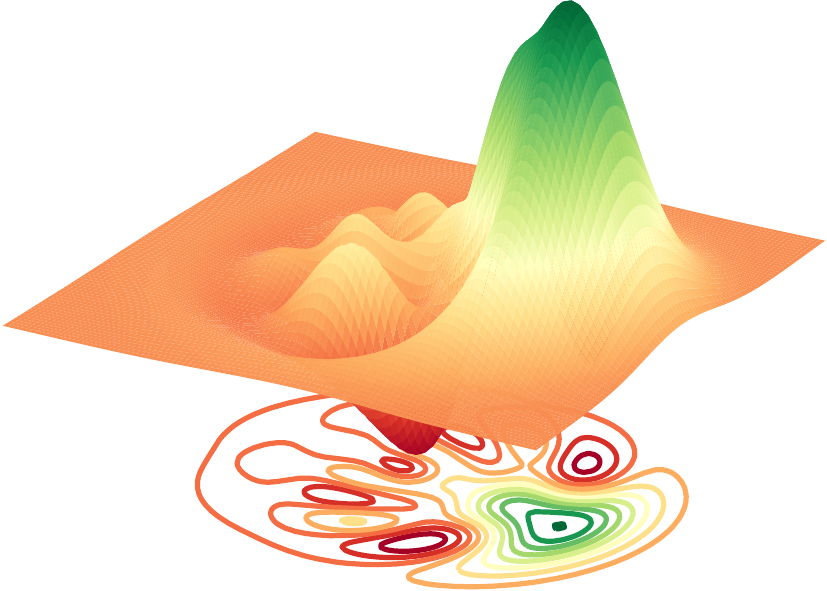}\\

\end{tabular}
\end{minipage}
\caption{(Top row) Visualizations of the transformation matrices for the ideal cubic phase gate (left) and the synthesized transformation $U(\vec{\theta})$ (right). In each case, the first (red) panel corresponds to the real part of the matrix and the second (green) panel to the imaginary part. Each square in the panel represents an element of the matrix in the Fock basis, with dark squares showing large positive values, light squares large negative values, and squares of neutral brightness corresponding to zero entries. The average gate fidelity between both transformations is 99.86\%. (Bottom row) Wigner functions of the states resulting from applying the ideal cubic phase gate (left) and the synthesized gate (right) to the equal superposition state $\ket{\Psi_{10}}$ of Eq.~(\ref{Eq:EqSup}).}\label{Fig:cubicphase}
\end{figure}
\end{center}

The cubic phase gate can increase photon number and therefore in general it maps states in the subspace of at most $d-1$ photons to states that may have more than $d-1$ photons. This transformation can be represented in terms of a matrix with $d$ columns where column $i$ represents the result of applying the gate to the $i$-th Fock state and the columns have dimension $d'>d$. The result of optimization is illustrated in Fig.~\ref{Fig:cubicphase}.  Here, we visualize the real and imaginary parts of the transformation matrices for both the ideal cubic phase gate and the unitary $U(\vec{\theta})$ applied by the circuit.

The synthesized gate has an average fidelity of $99.86 \%$ with respect to the ideal cubic phase gate. To visualize these transformations, we also highlight the result of applying both the ideal and synthesized gate to the equal superposition state in the $d$-dimensional subspace given by
\begin{equation}\label{Eq:EqSup}
\ket{\Psi_{d}} = \frac{1}{\sqrt{d}} \sum_{i=0}^{d-1} \ket{i},
\end{equation}
with $\ket{i}$ the $i$-th Fock state. In this case, we apply the ideal and synthesized gates to the state $\ket{\Psi_{10}}$ and plot the Wigner functions of the resultant states in Fig.~\ref{Fig:cubicphase}.

\begin{center}
\begin{figure}[t!]
\begin{tabular}{cc}
\includegraphics[width=1\columnwidth]{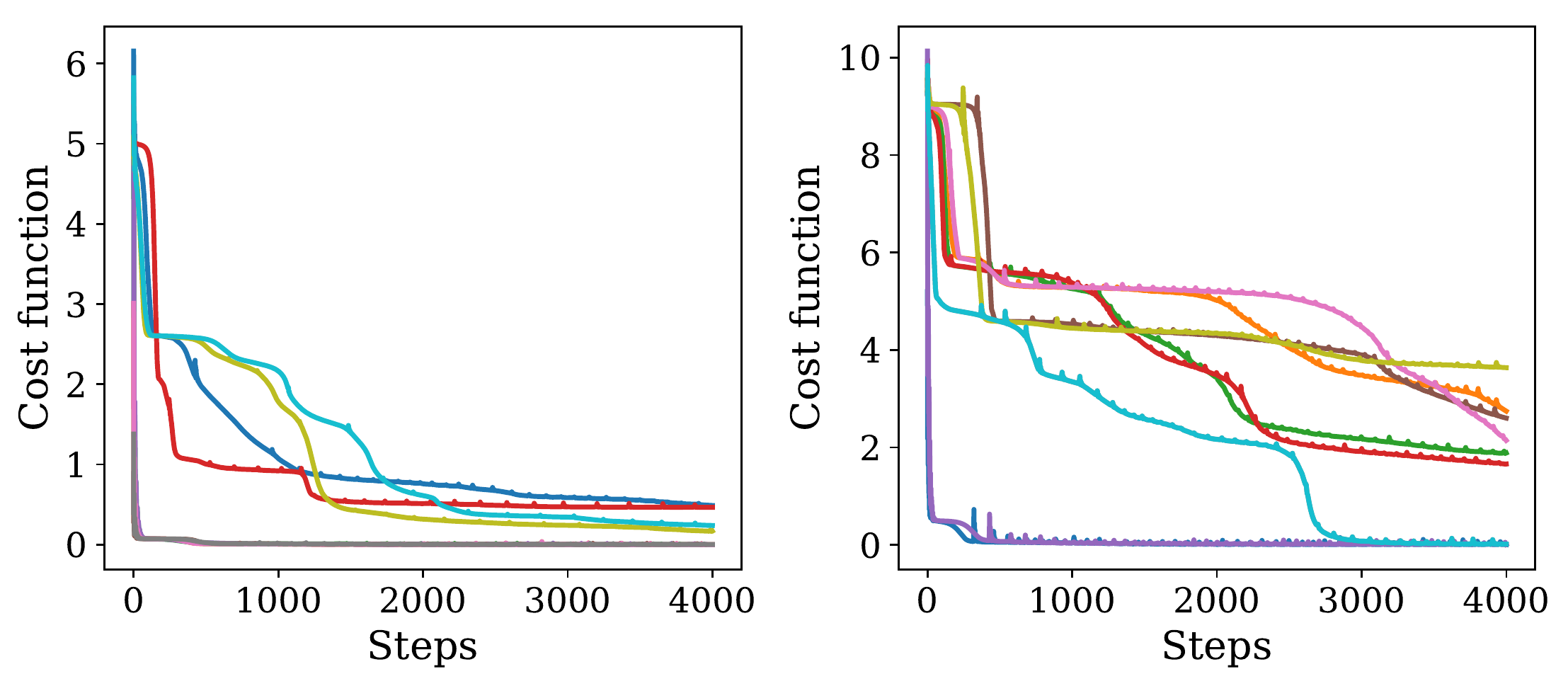}
\end{tabular}
\caption{Progress in minimizing the cost function for 10 independent runs of the optimization algorithm for the cubic phase gate  with parameter $\gamma=0.01$ acting on a $6$-dimensional subspace (left) and a $10$-dimensional subspace (right).}\label{Fig:Gatetrajectories}
\end{figure}
\end{center}

Similarly to the state preparation setting, the result of gate synthesis is non-deterministic due to random initialization of gate parameters and the use of stochastic gradient descent. In Fig.~\ref{Fig:Gatetrajectories}, we plot the progress in minimizing the cost function for $10$ independent runs of gate synthesis. We study both the case of a $6$-dimensional and a 10-dimensional subspace. The result of different optimization runs varies greatly, with many of them becoming stuck in local minima that are far from optimal. This behavior can be understood in terms of the subspace dimension: increasing the number of relations that define the transformation makes the optimization landscape more complex and adds computational overhead. In particular, it can be seen in Fig.~\ref{Fig:Gatetrajectories} that for $d=6$ input-output relations, a smaller proportion of runs become stuck in local minima that are far from the global optimum.

\subsubsection{Quantum Fourier transform}
\begin{center}
\begin{figure}[t!]
\begin{minipage}{0.22\columnwidth}
\includegraphics[width=\columnwidth]{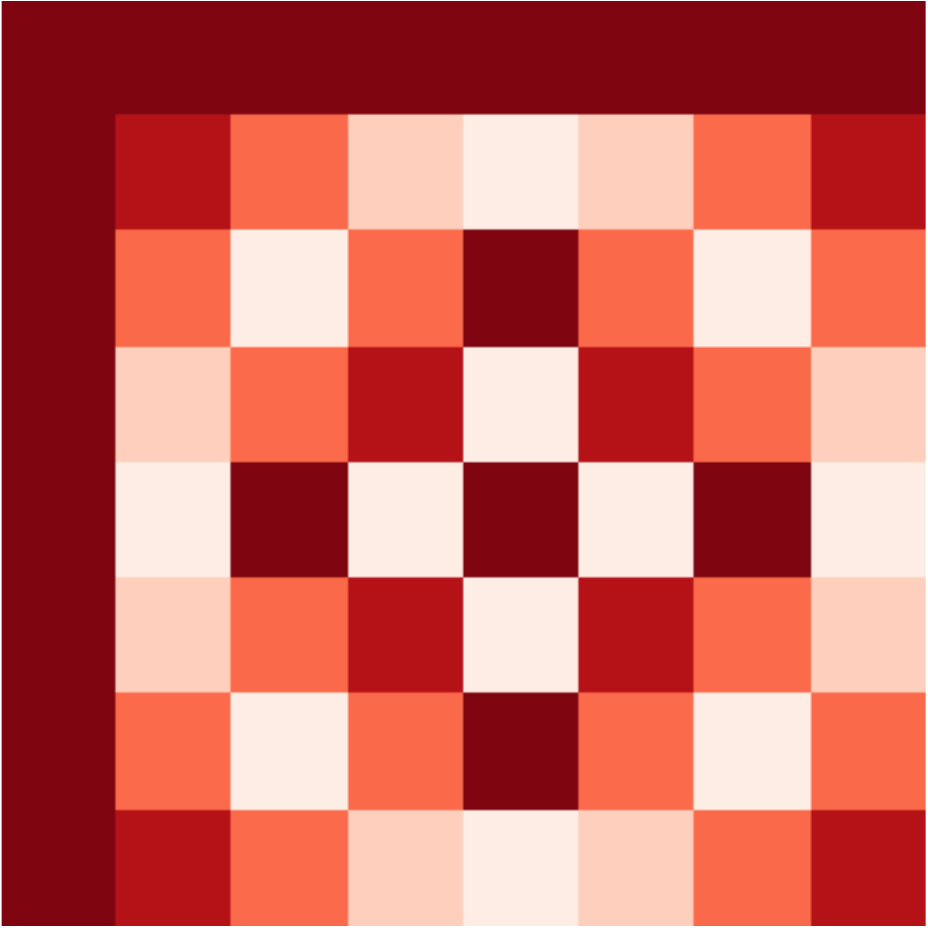}
\end{minipage}
\hspace{-4pt}
\begin{minipage}{0.22\columnwidth}
\includegraphics[width=\columnwidth]{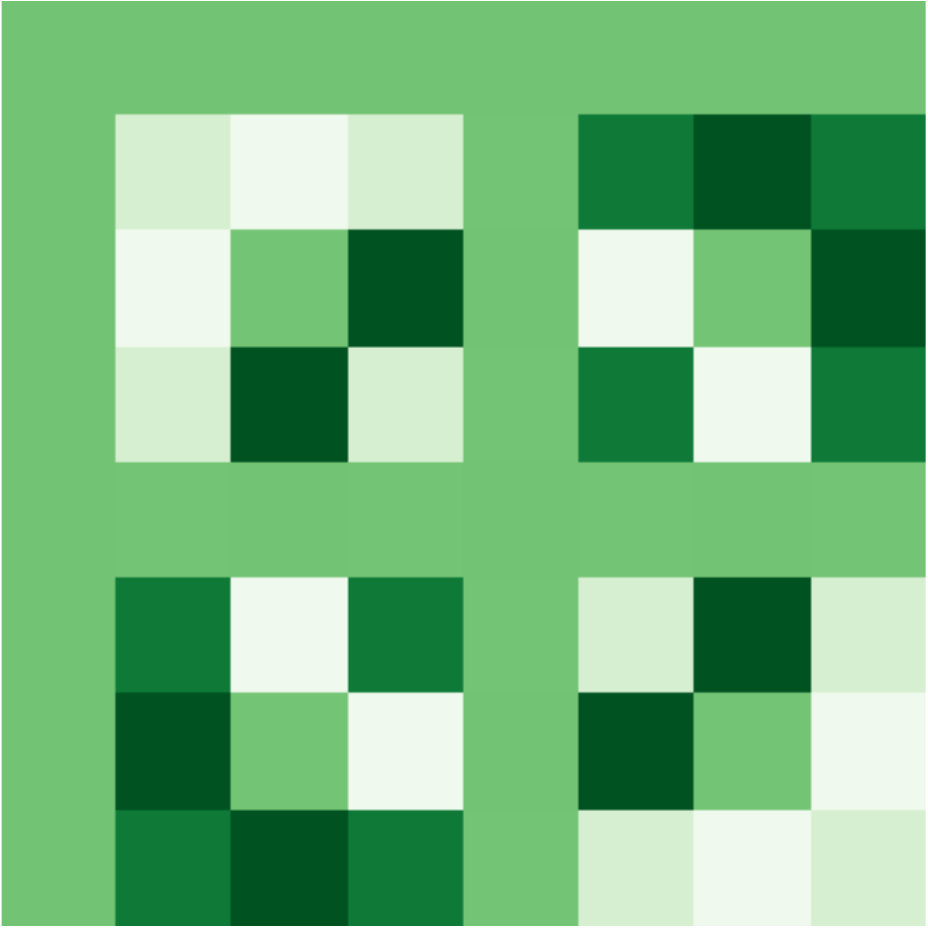}
\end{minipage}
\hspace{5pt}
\begin{minipage}{0.22\columnwidth}
\includegraphics[width=\columnwidth]{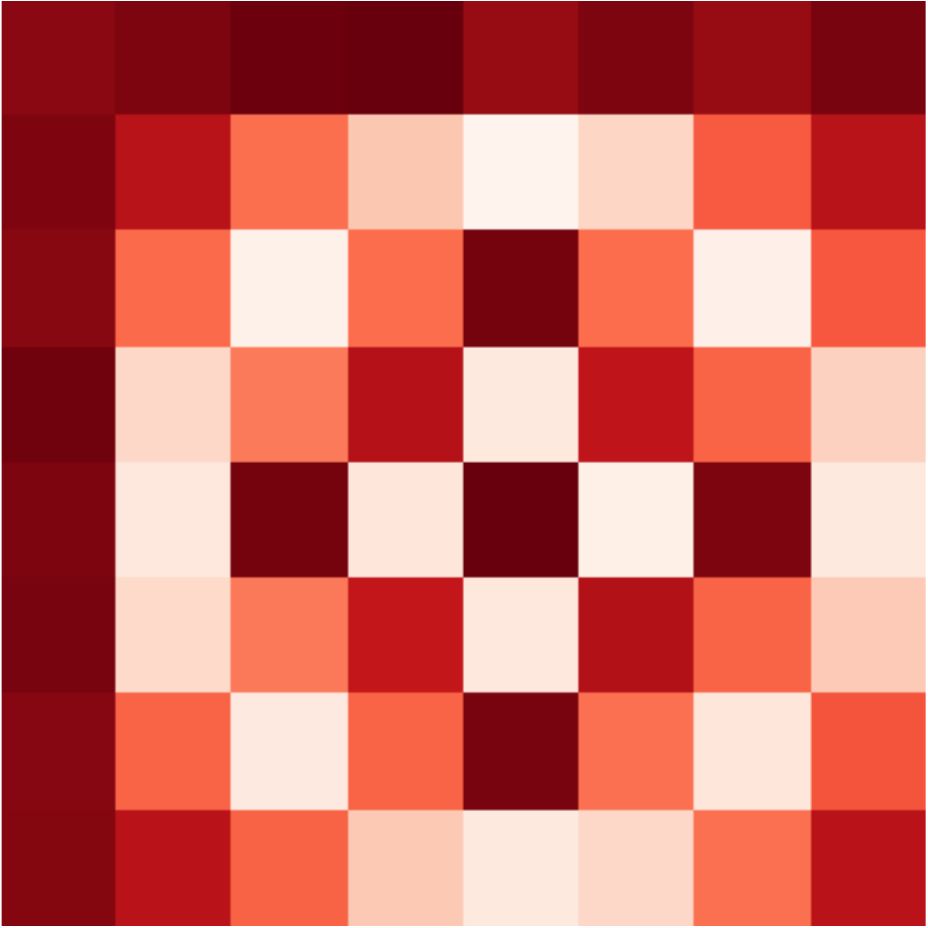}
\end{minipage}
\hspace{-4pt}
\begin{minipage}{0.22\columnwidth}
\includegraphics[width=\columnwidth]{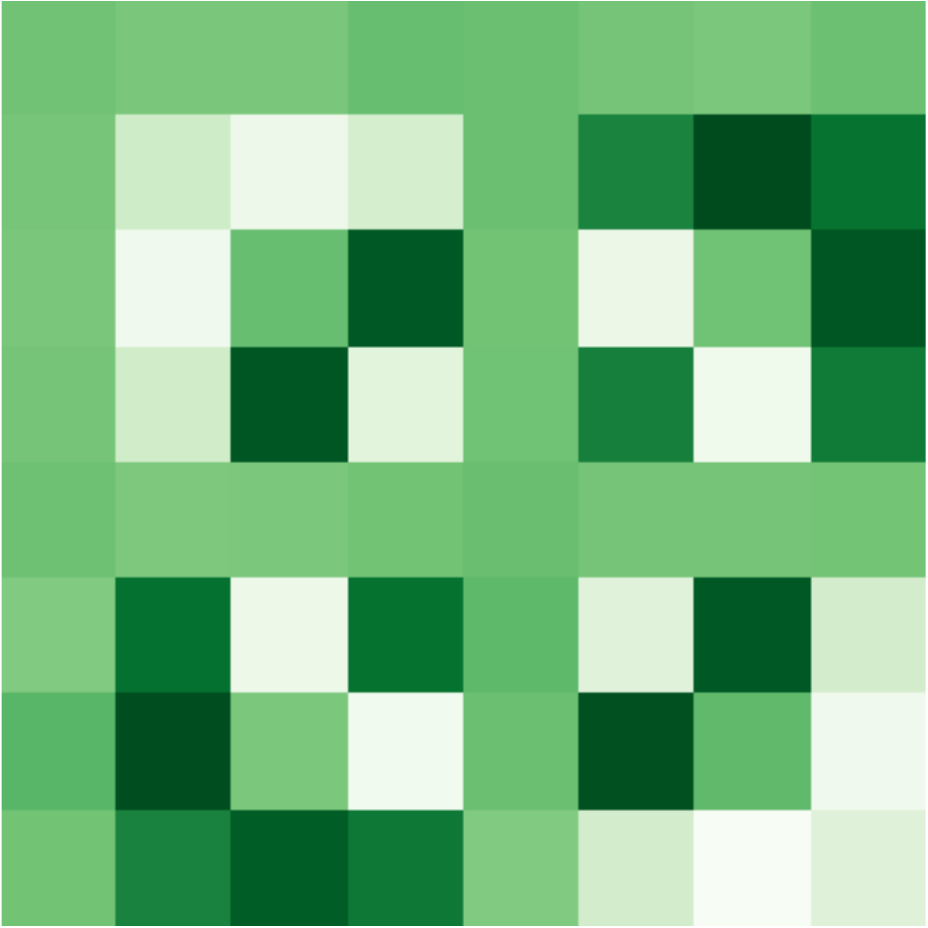}
\end{minipage}
\begin{minipage}{0.98\columnwidth}
\vspace{0.5cm}
\begin{tabular}{cc}
\includegraphics[width=0.48\columnwidth]{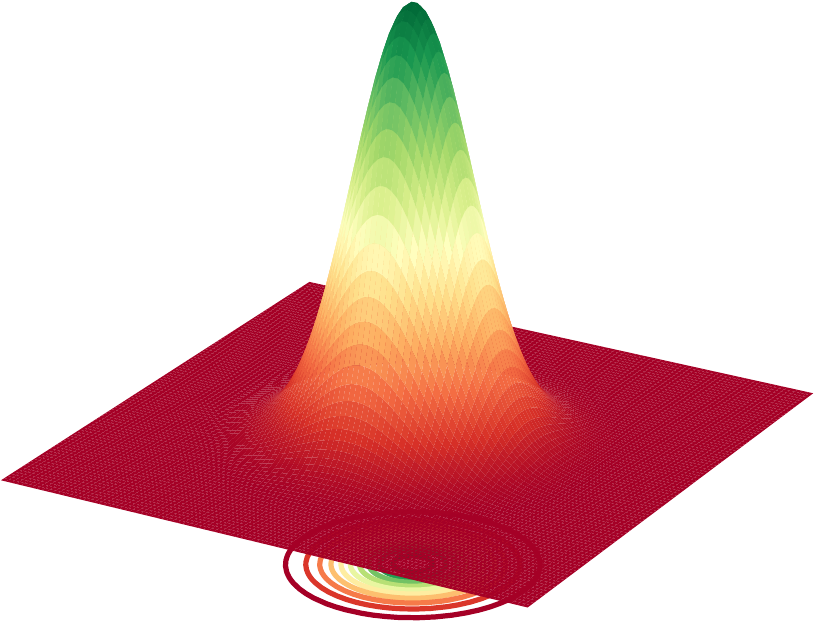}&
\includegraphics[width=0.48\columnwidth]{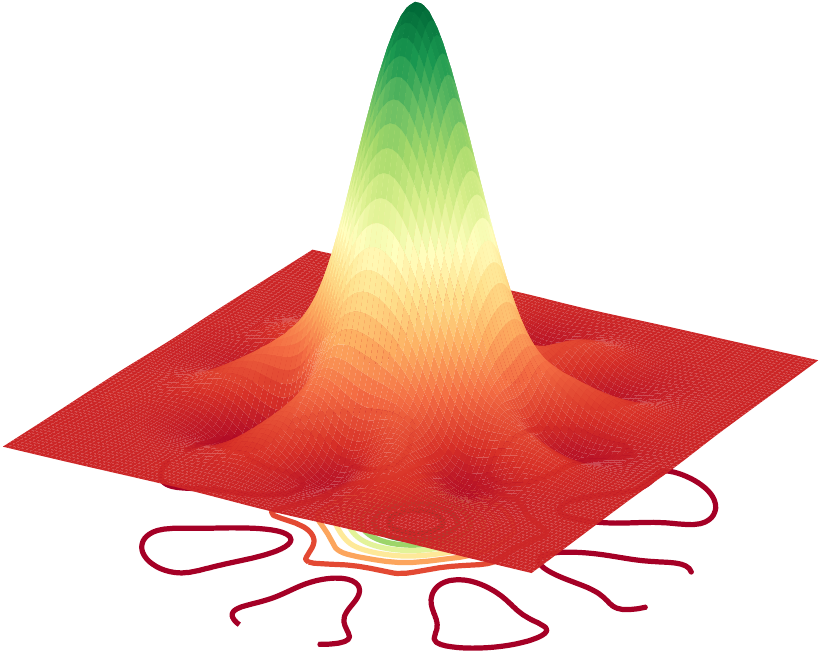}\\

\end{tabular}
\end{minipage}
\caption{(Top row) Visualizations of the transformation matrices for the ideal quantum Fourier transform unitary acting on the Fock basis (left) and the synthesized transformation $U(\vec{\theta})$ (right). In each case, the first (red) panel corresponds to the real part of the matrix and the second (green) panel to the imaginary part. Each square in the panel represents an element of the matrix in the Fock basis, with dark squares showing large positive values, light squares large negative values, and squares of neutral brightness corresponding to zero entries. The average gate fidelity between both transformations is 98.89\%. (Bottom row) Wigner functions of the states resulting from applying the ideal random gate (left) and the synthesized gate (right) to the equal superposition state $\ket{\Psi_{8}}$ of Eq.~(\ref{Eq:EqSup}). Note that the ideal resultant state is the vacuum.}\label{Fig:qft}
\end{figure}
\end{center}

We now consider a unitary transformation acting on a finite-dimensional system embedded into the first $d$ Fock states of a single mode. This transformation has the effect of mapping input Fock states into equal superpositions of all Fock states, thus performing a transformation between two mutually-unbiased bases. We denote this gate as the quantum Fourier transform (QFT) in relation to the mathematically equivalent transformation employed in discrete systems. The gate is defined as  
\beq
V_{\text{QFT}}= \frac{1}{d}\sum_{n=0}^{d-1}\sum_{m=0}^{d-1}e^{ \frac{2\pi i}{d} mn}\ket{n}\bra{m}.
\eeq
This is a highly complex gate and hence we allow for a deeper architecture and more optimization steps. We fix $d=8$ and carry out gate synthesis with $40$ layers (200 gates) for $8000$ steps, resulting in an average gate fidelity of $98.89 \%$ between the ideal gate and the transformation performed by the network. We plot visualizations of both unitaries in Fig.~\ref{Fig:qft} as well as the result of applying them to the equal superposition state $\ket{\Psi_{8}}$. Note that in this case $V_{\text{QFT}}\ket{\Psi_{8}}=\ket{0}$.

\subsubsection{Random unitary}

\begin{center}
\begin{figure}[t!]
\begin{minipage}{0.22\columnwidth}
\includegraphics[width=\columnwidth]{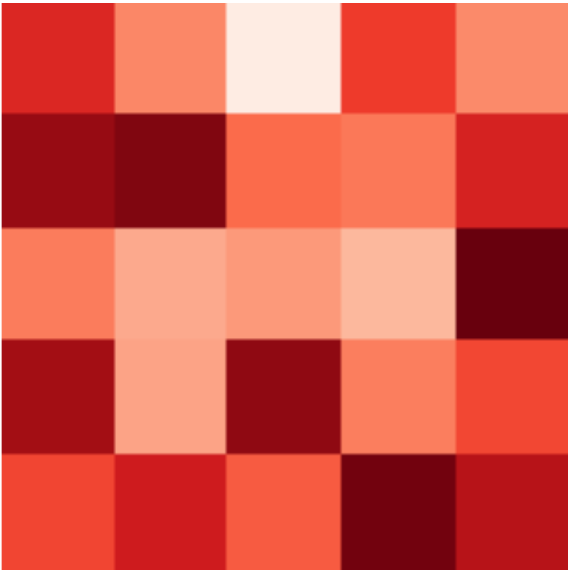}
\end{minipage}
\hspace{-4pt}
\begin{minipage}{0.22\columnwidth}
\includegraphics[width=\columnwidth]{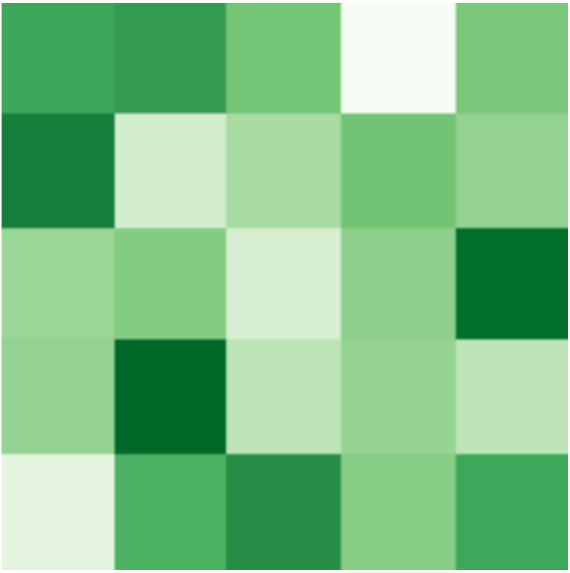}
\end{minipage}
\hspace{5pt}
\begin{minipage}{0.22\columnwidth}
\includegraphics[width=\columnwidth]{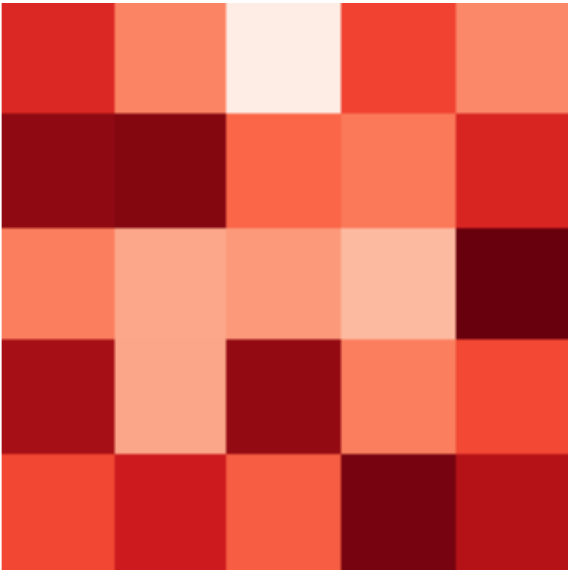}
\end{minipage}
\hspace{-4pt}
\begin{minipage}{0.22\columnwidth}
\includegraphics[width=\columnwidth]{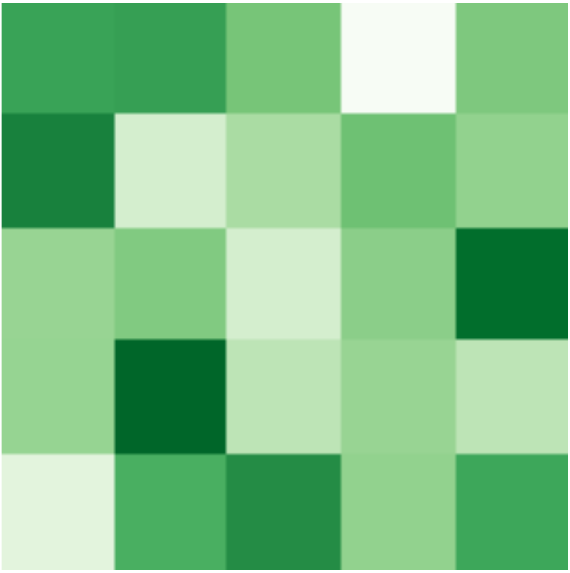}
\end{minipage}
\begin{minipage}{0.98\columnwidth}
\vspace{0.5cm}
\begin{tabular}{cc}
\includegraphics[width=0.48\columnwidth]{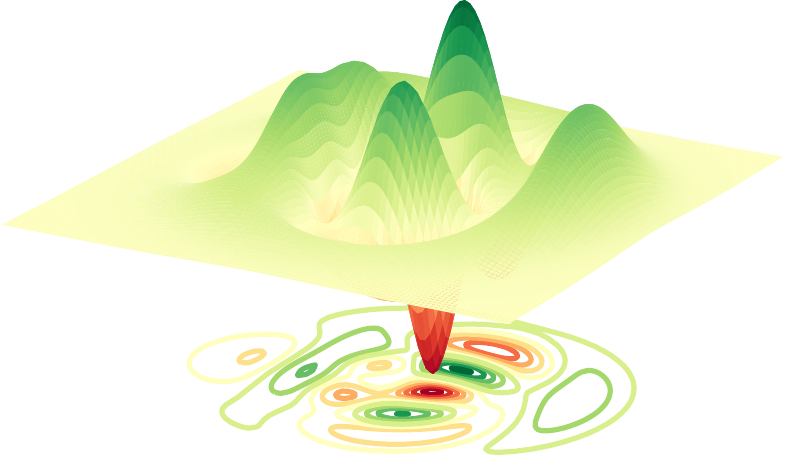}&
\includegraphics[width=0.48\columnwidth]{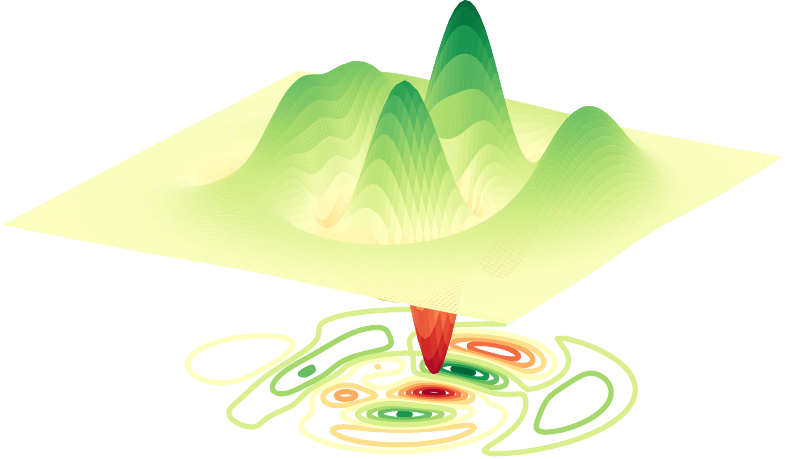}\\

\end{tabular}
\end{minipage}
\caption{(Top row) Visualizations of the transformation matrices for the ideal random gate (left) and the synthesized transformation $U(\vec{\theta})$ (right). In each case, the first (red) panel corresponds to the real part of the matrix and the second (green) panel to the imaginary part. Each square in the panel represents an element of the matrix in the Fock basis, with dark squares showing large positive values, light squares large negative values, and squares of neutral brightness corresponding to zero entries. The average gate fidelity between both transformations is 99.50\%. (Bottom row) Wigner functions of the states resulting from applying the ideal random gate (left) and the synthesized gate (right) to the equal superposition state $\ket{\Psi_{5}}$. }\label{Fig:randomu}
\end{figure}
\end{center}

As in the case of state preparation, we now study the suitability of our gate synthesis method for unstructured transformations. This complements our investigation of random state preparation by adding multiple relations. For this task, we generate a random unitary $V$ according to the Haar measure that acts only on the five-dimensional space of at most four photons. Gate synthesis was performed with $25$ layers (125 gates) for $1000$ steps, resulting in an average fidelity of $99.5\%$. Along with the previous QFT gate, this shows the viability of our gate synthesis method for complex transformations. Figure~\ref{Fig:randomu} visualizes the target and learned unitaries, along with the Wigner functions resulting from applying these unitaries to the $d=5$ superposition state $\ket{\Psi_{5}}$.

\begin{table*}[t!]
	\centering
	\begin{ruledtabular}
		\begin{tabular}{lcccc}
			\textit{Hyperparameters }&\textit{Cubic phase} & \textit{QFT in the Fock basis} & \textit{Random unitary} &\textit{Cross-Kerr}\\\hline
			Fock basis cutoff dimension $D$ & 20 & 18 & 16 & 9\\
			Number of input-output relations  & 10 & 8 & 5 & 25\\
			Depth/number of layers & 25 & 40 & 25 & 25\\
			Total elementary gate count & 125 & 200 & 125 & 125\\
			Number of optimization steps & 20000 & 8000 & 10000 & 10000\\
			Target gate parameter & $\gamma=0.1$ & - & - & $\kappa=0.1$\\\hline
			\textit{Results and circuit parameters} & & & &\\\hline
			Average fidelity to the target gate $\bar{F}$ & $99.86\%$ & $98.89\%$ & $99.50\%$ & $99.994\%$ \\
			Runtime (seconds)  & 1,966 & 17,125 & 2,631 & 81,935\\
			Maximum absolute displacement $|\alpha|$ & $0.2113$ & $0.5278$ &$0.7615$ & $5.135\times10^{-5}$ \\
			Maximum absolute squeezing $|r|$ & $0.1242$ & $0.2543$ &$0.2753$ & $5.537\times 10^{-5}$ \\
			Maximum absolute Kerr strength $|\kappa|$ & $0.02437$ & $0.08323$ & $0.3428$ & $0.01726$ 
		\end{tabular}
	\end{ruledtabular}
Optimization was run on a quadcore AMD Radeon R7 CPU operating at 2.1GHz with 12GB RAM except for the cubic phase gate and the QFT gate, which where optimized employing a 20-core Intel Xeon CPU operating at 2.4GHz with 252GB RAM. 
	\caption{Summary of gate synthesis hyperparameters and learning results.}\label{tab:gateparameters}
	
\end{table*}

\subsubsection{Cross-Kerr interaction}

We now extend our investigation of gate synthesis to two-mode unitaries. We focus on the cross-Kerr interaction, which mediates coupling between photons in different modes by applying a phase that depends on the photon number in each mode. It can be written as
\beq
V_{\text{CK}}(\kappa)=\exp \left(-i \kappa \hat{n}_{1} \hat{n}_{2}\right),
\eeq
with $\hat{n}_{k}$ the number operator on mode $k$ and $\kappa$ the interaction strength. The cross-Kerr is a widely used non-linear bosonic interaction, appearing in the context of superconducting circuits and quantum photonics. It has been shown to allow the implementation of a nearly-deterministic qubit CNOT gate with vastly reduced resources compared to other approaches \cite{nemoto2004nearly}, and leads to a maximally efficient implementation of the non-linear sign gate on photonic qubits \cite{van2011optical}. The cross-Kerr interaction is also vital in state preparation, for instance in the generation of arbitrary-strength Schr\"odinger cat states \cite{vitali1997conditional,clausen2002quantum}, and represents the nearest-neighbor or dipole interaction in the Bose-Hubbard model \cite{kalajdzievski2018continuous}.

\begin{center}
\begin{figure}[t!]
\begin{minipage}{0.22\columnwidth}
\includegraphics[width=\columnwidth]{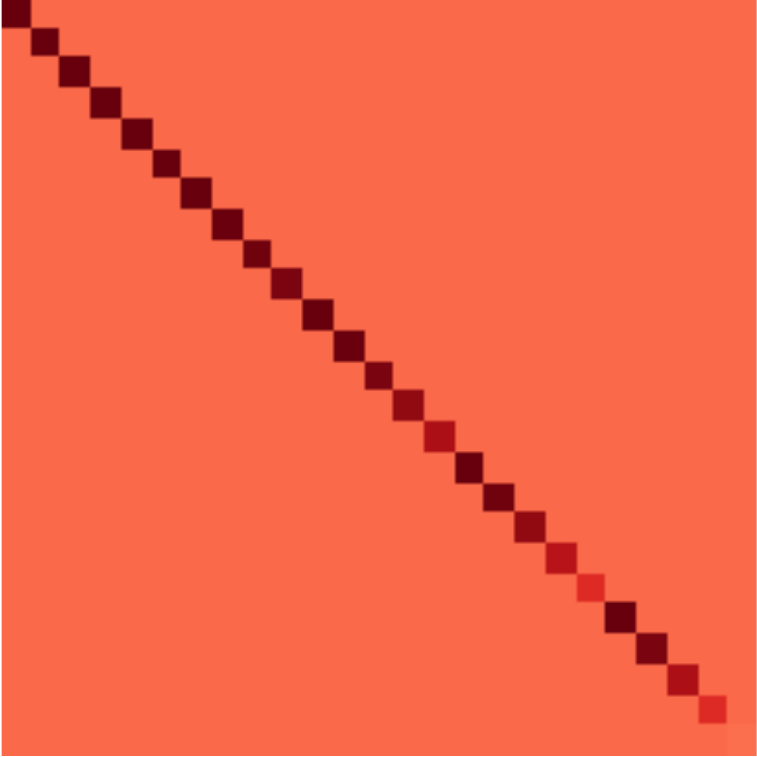}
\end{minipage}
\hspace{-4pt}
\begin{minipage}{0.22\columnwidth}
\includegraphics[width=\columnwidth]{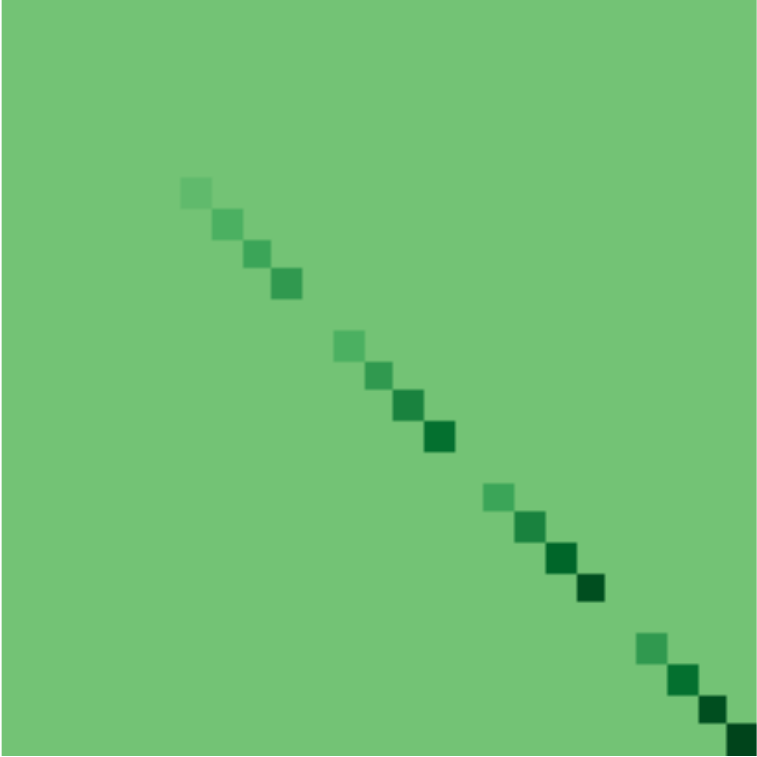}
\end{minipage}
\hspace{5pt}
\begin{minipage}{0.22\columnwidth}
\includegraphics[width=\columnwidth]{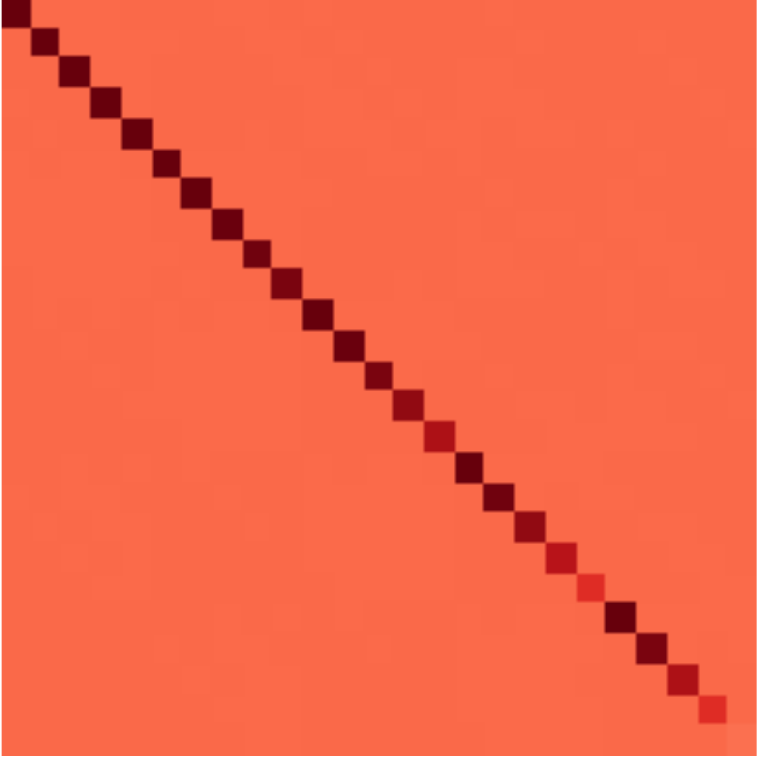}
\end{minipage}
\hspace{-4pt}
\begin{minipage}{0.22\columnwidth}
\includegraphics[width=\columnwidth]{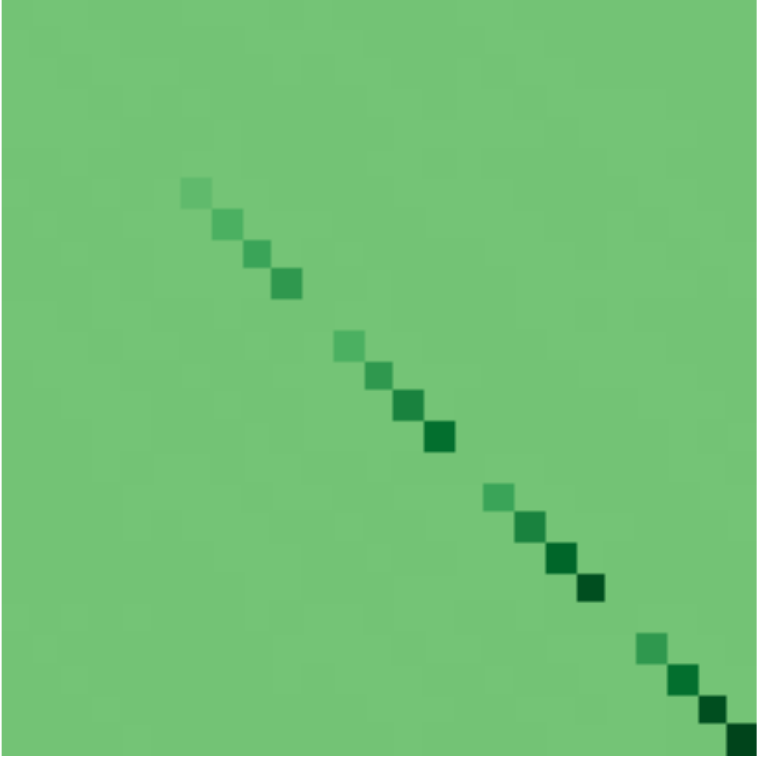}
\end{minipage}
\begin{minipage}{0.98\columnwidth}
\vspace{0.5cm}
\begin{tabular}{cc}
\includegraphics[width=0.48\columnwidth]{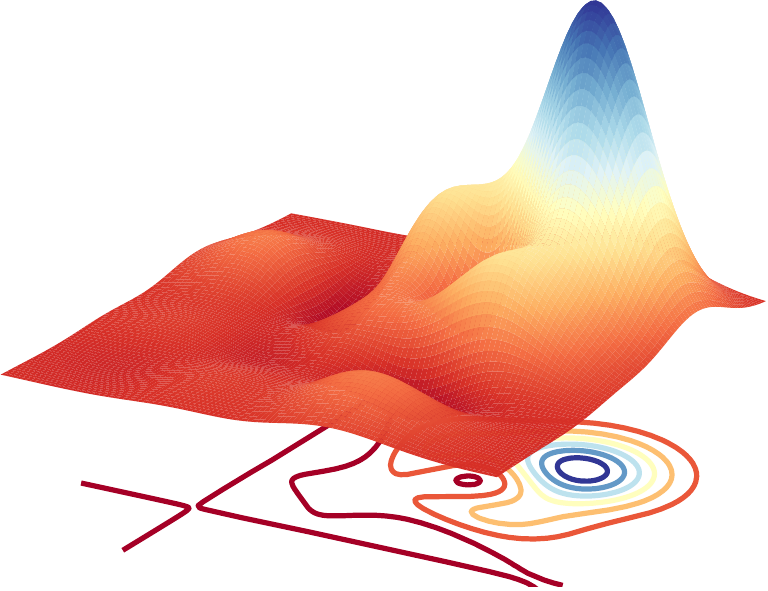}&
\includegraphics[width=0.48\columnwidth]{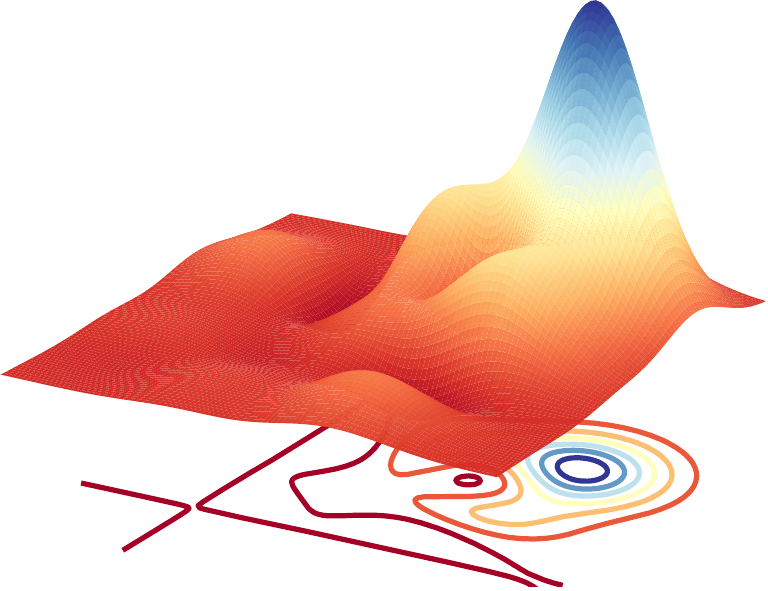}\\

\end{tabular}
\end{minipage}
\caption{(Top row) Visualizations of the transformation matrices for the ideal two mode cross-Kerr gate (left) and the synthesized transformation $U(\vec{\theta})$ (right). In each case, the first (red) panel corresponds to the real part of the matrix and the second (green) panel to the imaginary part. Each square in the panel represents an element of the matrix in the Fock basis, with dark squares showing large positive values, light squares large negative values, and squares of neutral brightness corresponding to zero entries. To represent a two-mode unitary as a matrix, we use the lexicographical ordering $(\ket{0,0},\dots, \ket{0,d-1}, \ket{1,0},\dots\dots,\ket{d-1,d-1})$ to label the rows and columns. The average gate fidelity between both transformations is 99.994\%.(Bottom row) Wave functions of the states resulting from applying the ideal two mode cross-Kerr gate (left) and the synthesized gate (right) to the equal superposition state $|\Psi_{5}^{(2)}\rangle$. }\label{Fig:crosskerr}
\end{figure}
\end{center}

We synthesize the cross-Kerr interaction using $25$ layers (125 gates) and $10000$ optimization steps, with $\kappa = 0.1$. By restricting to the first $d=5$ Fock states in each mode, our architecture can approximate the cross-Kerr interaction with an average fidelity to the ideal unitary of $99.994 \%$. The result of cross-Kerr gate synthesis is visualized in Fig.~\ref{Fig:crosskerr}. The cross-Kerr optimization resulted in a circuit with negligible levels of displacement and squeezing, confirming the intuition that the use of Kerr gates and beamsplitters is sufficient in this case.

We test the action of both ideal and learnt unitaries by applying them to the equal superposition two-mode state, which is written as
\begin{equation}
|\Psi_{d}^{(2)}\rangle = \frac{1}{d} \sum_{i=0}^{d-1}\sum_{j=0}^{d-1} \ket{i}\ket{j}.
\end{equation}
To visualize the resultant states, we plot the two-dimensional wavefunctions instead of the Wigner functions, similarly to two-mode state preparation.

Extending to the two-mode setting increases the computational overhead of gate synthesis. Indeed, as in state preparation, the truncated Hilbert space of the composite system is quadratically larger. However, an additional overhead in gate synthesis is that the number of relations that define the transformation also increases for two modes, e.g., to $25$ for $d=5$. Nevertheless, our results demonstrate synthesis of high fidelity approximations of a cross-Kerr gate with parameter $\kappa = 0.1$. Historically, the interaction strength for the cross-Kerr gate that is realizable with available technology has been too small for use in quantum computing. Recent results have considered decompositions of the cross-Kerr into cubic phase gates, controlled-phase gates, and beamsplitters ~\cite{sefi2011decompose,douce2018probabilistic}, requiring approximately 1000 elementary operations for a decomposition precision of $\sim0.1$. Our approach uncovers a short-depth decomposition, requiring only $25$ Kerr gates and $125$ elementary operations in total. A summary of the gate synthesis results is reported in Table \ref{tab:gateparameters}.

\section{Conclusion}
We have presented a general method to leverage techniques from machine learning and optimization to find quantum circuits that are capable of reproducing desired transformations between input and output states. Our results make it possible to largely automate the task of discovering circuits that perform specific subroutines of quantum algorithms. Crucially, we are capable of synthesizing high-fidelity states and gates using short-depth circuits -- a feature that makes our techniques particularly well suited for near-term quantum devices.

The strategies employed here -- namely stochastic gradient descent based on automatic differentiation of simulated quantum circuits -- can be straightforwardly applied to other forms of quantum computation. However, despite the usefulness of automatic differentiation for quantum computing and quantum simulation \cite{steiger2005using,leung2017speedup, tamayo2018automatic}, to our knowledge, Strawberry Fields is currently the only quantum software library which natively supports automatic differentiation. Further work is required to bring this feature to other simulation libraries.

Finally, it is important to extend these methods to more general quantum algorithms involving complex transformations between several modes. The techniques outlined in this work require the ability to classically simulate circuits, a task that becomes intractable as the number of modes increases. Achieving this extension will likely require the use of specialized quantum techniques to optimize the circuits directly.

\acknowledgements We thank Krishna Kumar Sabapathy, Pierre-Luc Dallaire-Demers, Maria Schuld, and Nicol\'as Quesada for valuable discussions. We also thank Nicolas Menicucci for suggesting Ref.~\cite{noh2018improved}.

\pagebreak

\bibliographystyle{apsrev}
\bibliography{Bibliography}

\appendix

\section{Simulation cutoff size}\label{Appendix:CutoffVsRelations}

The numerical simulations require a truncation in the Fock basis to a fixed cutoff  dimension $D$. This means that any state $\ket{\Psi}$ is described in the simulations by $\Pi_{D}\ket{\Psi}$, where $\Pi_{D}$ is a projector onto the $D$-dimensional truncated Hilbert space. To ensure reliable simulations, it is important to fix a suitably large value of $D$.
 
When synthesizing states or gates with support only in a restricted Hilbert space, we set $D$ to be approximately five steps larger than the dimension of the restricted space, allowing for some leeway for the gates used in each layer to map internally out of the space at certain points of the circuit. Additionally, for synthesizing gates, $D$ is chosen so that the target states remain close to being normalized when calculated in the truncated space. Specifically, the output states $|\Psi_{t}^{(i)}\rangle$ of the ideal unitary for any $i$ of $d$ input-output relations ($d=1$ for state preparation) should satisfy
\begin{equation}\label{Eq:CutoffEquation}
\| \Pi_{D}|\Psi_{t}^{(i)}\rangle \|_2 \geq 1- \epsilon ,
\end{equation}
where $\| \cdot \|_2$ is the $2$-norm and $\epsilon$ is a fixed tolerance. Note that increasing $D$ significantly adds computational overhead. In the relevant situations presented here, we set $D$ to be the smallest value such that Eq.~\eqref{Eq:CutoffEquation} is satisfied with $\epsilon = 0.0001$.

\section{Fidelity for gate synthesis}\label{Appendix:AvgMinFidelity}
The average fidelity of two quantum unitary operations $U$ and $V$ is defined by \cite{nielsen2002simple}
\begin{align}
	\bar{F} = \int d\Psi \langle \Psi \mid V^\dagger U\ket{\Psi}\bra{\Psi}U^\dagger V \mid \Psi \rangle
\end{align}
where the integral is performed over the Haar measure on the state space.

To calculate the average fidelity, we employ another common fidelity measure known as the process fidelity. The process fidelity is calculated by first applying the $N$-mode unitary $V$ defined on the $d$-dimensional Fock subspace to one half of a maximally entangled state $\ket{\phi}$,
\begin{align}
	\ket{\Psi(V)} =(\id\otimes V)\frac{1}{\sqrt{d}}\sum_j\ket{jj}.
\end{align}	
This is repeated for the learnt unitary $U(\vec{\theta})$, and the process fidelity determined by taking the corresponding state overlap:
\begin{align}
F_{\epsilon} = |\langle{\Psi(V)}\mid{\Psi(U(\vec{\theta}))}\rangle|^2.
\end{align}
The process fidelity is advantageous, as it is bounded by the average fidelity of only two complementary sets of input states, avoiding the need to sample over the entire state space \cite{hofmann2005complementary,reich2013minimum}. Furthermore, the process fidelity is related directly to the average fidelity via the relationship \cite{horodecki1999general}
\begin{align}\label{eq:process}
	\bar{F} = \frac{F_{\epsilon}d+1}{d+1}.
\end{align}

It should be noted that a slight subtlety arises in the case of computing the process fidelity for the cubic phase gate. While $U_{CP}(\vec{\theta})$ acts on a $d$-dimensional subspace, it has the potential to map states within this subspace into the larger $D$-dimensional subspace given by the Fock basis cutoff dimension. As a result, to numerically approximate the average gate fidelity in this case, we compute the quantity
\begin{align}\label{eq:avgfid}
	&\bar{F}  = \frac{1}{N} \sum_{i=0}^{N} |\langle 0 \mid W_i^\dagger V^\dagger U(\vec{\theta}) W_i \mid 0\rangle|^2,
\end{align}
where $W_i\in\{W_1,\dots,W_N\}$, a set of Haar-distributed $d\times d$ unitary matrices. For the results presented in \autoref{sec:gate}, we calculate this quantity using an ensemble of $N=10^4$ randomly chosen matrices from the Haar measure.

\end{document}